\documentclass[11pt,showpacs,preprintnumbers,amsmath,amssymb]{revtex4}
\usepackage{amsmath,amssymb,graphics}
\usepackage{epsfig}
\newcommand{\nc}{\newcommand}
\nc{\ba}{\begin{eqnarray}} \nc{\ea}{\end{eqnarray}}
\newcommand\be{\begin{equation}}
\newcommand\ee{\end{equation}}

\usepackage{epstopdf}

\nc{\x}{{\bf{x}}}
\nc{\e}{{\bf{e}}}

\input{epsf.sty}

\preprint{IPM/P-2010/043 }

\begin{document}

\title{Anisotropic Inflation from Charged Scalar Fields }
\author{Razieh Emami$^{1}$}
\email{emami(AT)ipm.ir}
\author{Hassan Firouzjahi$^{1}$}
\email{firouz(AT)ipm.ir}
\author{S. M. Sadegh Movahed$^{2}$}
\email{m.s.movahed(AT)ipm.ir}
\author{Moslem Zarei$^{3}$}
\email{m.zarei(AT)cc.iut.ac.ir}
\affiliation{1) School of Physics, Institute for
Research in Fundamental Sciences (IPM), P. O. Box 19395-5531,
Tehran, Iran}
\affiliation{2)Department of Physics, Shahid Beheshti University,
G.C., Evin, Tehran 19839, Iran }
\affiliation{3)Department of Physics, Isfahan University of Technology, Isfahan 84156-83111, Iran }

\begin{abstract}

We consider models of inflation with $U(1)$ gauge fields and charged scalar fields including symmetry breaking potential, chaotic inflation and hybrid inflation.
We show that there exist attractor solutions where the anisotropies produced during inflation 
becomes comparable to the slow-roll parameters.  
In the models where the inflaton field is a charged scalar field  the gauge field becomes highly oscillatory at the end of inflation ending inflation quickly. Furthermore, in charged hybrid inflation the onset of waterfall phase transition at the end of inflation is affected significantly by the evolution of the background gauge field. 
Rapid oscillations of the gauge field and its coupling to inflaton can have interesting effects 
on preheating and non-Gaussianities. 

\end{abstract}

\maketitle

\newpage


\section{Introduction}

Cosmic Inflation proved to be a successful theory of early universe and the mechanism of  structure formation \cite{Komatsu:2010fb}. The simplest modes of inflation are based on a scalar field coupled minimally to gravity. The scalar field potential is flat enough to allow for about 60 e-foldings or so to solve the flatness and the horizon problem of the standard big-bang cosmology. Precision data can distinguish amongst different inflationary scenarios based on their predictions for the spectral index of curvature perturbation, the amplitude of gravitational wave or the degrees of non-Gaussianities induced on cosmological perturbations.

There have been considerable interests on primordial anisotropies both observationally and theoretically. Observationally, there may be some indications of the the statistical anisotropy of the comic microwave background (CMB)  \cite{Eriksen:2003db} although the statistical significances of these findings are under debate \cite{Komatsu:2010fb, Hanson:2009gu, Hanson:2010gu}. On the theoretical sides there have been many attempts to construct models of inflation with vector fields or gauge fields
which can create sizable amount of anisotropy on curvature perturbations 
\cite{Ford:1989me, Kaloper:1991rw,  Kawai:1998bn,  Barrow:2005qv, Barrow:2009gx, Campanelli:2009tk, Golovnev:2008cf, Kanno:2008gn, Pitrou:2008gk, Moniz:2010cm, Boehmer:2007ut, Koivisto:2008xf}. These mechanisms can provide a seed of anisotropies at the order of few percent which may be detectable on CMB \cite{Ackerman:2007nb}, \cite{Yokoyama:2008xw}, \cite{Dimopoulos:2009vu, Dimastrogiovanni:2010sm, ValenzuelaToledo:2009af}.

There have been different approaches to implements vector field in models of inflation where the vector field breaks the gauge symmetry explicitly.  One fundamental problem in these models, as demonstrated in \cite{Himmetoglu:2008zp}, is the appearance of ghost which render the system unstable and physically unacceptable. Therefore, it is crucial that the vector field is protected by a gauge symmetry so the longitudinal mode of the vector field excitations is not physical. On the other hand, because of the conformal invariance of models with gauge fields, any excitation of gauge field during inflation is diluted and can not seed the desired anisotropies. Therefor it is essential that one breaks the conformal invariance while keeping the gauge symmetry explicit. This approach was employed in different contexts in 
\cite{Martin:2007ue, Watanabe:2009ct, Emami:2009vd, Gumrukcuoglu:2010yc, Dulaney:2010sq, Kanno:2009ei, Watanabe:2010fh, Dimopoulos:2010xq}.

In this paper we would like to study different inflationary models where there is a non-zero background  $U(1)$
gauge field, $A_\mu$, coupled to a complex scalar field. The charged scalar field can be either the inflaton field or the waterfall field of hybrid inflation. Furthermore, in order to break the conformal invariance and produce large enough anisotropies, as explained above, we assume that the gauge field has a time-dependent gauge kinetic coupling with the kinetic energy in the form of $\frac{-f(\phi)}{4 } F_{\mu \nu} F^{\mu \nu}$. To be specific, we consider three following models:

\begin{enumerate}

\item  Symmetry breaking hilltop inflation: In this model  inflaton is a charged complex scalar field with the symmetry breaking (Mexican hat) potential 
$ V = \frac{\lambda}{4 }  \left(  |\phi|^2 - \frac{M^2}{\lambda} \right)^2$. We work with the gauge kinetic coupling $f(\phi) = \left(\frac{\mu}{|\phi|} \right)^{p}$ with the appropriate parameter $p$. In the presence of a $U(1)$ gauge field, this symmetry breaking potential is theoretically well motivated.

\item  Charged hybrid inflation: This is the standard hybrid inflation \cite{Linde:1993cn} where the inflaton field $\phi$ is real but now the waterfall field $\psi$ is charged under $U(1)$ gauge field. The potential is 
 $ V = \frac{\lambda}{4 }  \left(  |\psi|^2 - \frac{M^2}{\lambda} \right)^2 + \frac{g^2}{2} \phi^2 |\psi|^2 + \frac{m^2}{2} \phi^2$
 and the gauge kinetic coupling is  $f(\phi) = \left(\frac{\phi}{\phi_c}\right)^p $ where $\phi_c\equiv M/g$ is the critical value of inflaton field at the time of waterfall phase transition.

\item Chaotic inflation with the potential $V= \frac{m^2}{2} |\phi|^2$ where the inflaton field $\phi$ is charged under the $U(1)$ field. The case with no gauge coupling  (real $\phi$) was studied in \cite{Watanabe:2009ct}. As in \cite{Watanabe:2009ct} we work with the gauge kinetic coupling $f(\phi) = e^{c |\phi|^2/2M_P^2}$ with the appropriate coupling $c$.
We generalize their results to our case with a non-zero gauge coupling. 

The rest of paper is organized as follows. In section \ref{background} we study our set up where the inflaton field is charged under the $U(1)$ gauge field with a time depending gauge kinetic coupling. In section \ref{symmetry} we study the symmetry breaking potential in details and calculate the level of anisotropy during inflation. In section \ref{hybrid} we concentrate on hybrid inflation. We will see that the presence of the gauge field plays important role at the end of inflation and dynamics of waterfall phase transition. In section \ref{chaotic} we briefly study the case of chaotic inflation where now the inflaton field is charged under the $U(1)$ field and  compare our results with those of \cite{Watanabe:2009ct}. Conclusion and brief discussions are given in section \ref{conclusion}.

\end{enumerate}

\section{Background Equations}
\label{background}
Here we present the action for the cases of symmetry breaking potential where the 
inflaton field $\phi$ is charged under the $U(1)$ gauge field with a $\phi$-dependent  gauge kinetic coupling $f^{2}(\phi)$. The action and the background equations for the other cases can be obtained accordingly.  

As in \cite{Emami:2009vd} the action is
 \ba
 \label{action} S= \int
d^4 x  \sqrt{-g} \left [ \frac{M_P^2}{2} R - \frac{1}{2} D_\mu \phi
\,  D^\mu \bar \phi -   \frac{f^{2}(\phi)}{4} F_{\mu \nu} F^{\mu
\nu}  - V(\phi, \bar \phi) \right]
\ea
where $M_P^{-2} = 8 \pi G$,
for $G$ being the Newton constant and the overline represents the
complex conjugation. The covariant derivative is given by
\ba
D_\mu
\phi = \partial_\mu  \phi + i \e \,  \phi  \, A_\mu
\ea
where $\e$ is
the dimensionless gauge coupling of $A_\mu$ to $\phi$. As usual, the
gauge field strength is given by
\ba F_{\mu \nu} = \nabla_\mu A_\nu
- \nabla_\nu A_\mu  = \partial_\mu A_\nu - \partial_\nu A_\mu \, .
\ea
We work with potentials which have axial symmetry where $V$ and
$f(\phi)$ are only functions of $\phi \bar \phi=  |\phi |^2$. It is
more instructive to decompose the inflaton field into the radial and
angular parts \ba \phi(x) = \rho(x) \,  e^{i \theta(x)}\, , \ea so
$V=V(\rho)$ and $f^2(\phi)=f^2(\rho)$. As usual, the action
(\ref{action}) is invariant under local gauge transformation \ba
\label{transformation} A_\mu \rightarrow A_\mu - \frac{1}{e}
\partial_\mu \epsilon(x) \quad , \quad \theta \rightarrow \theta +
\epsilon(x) \, . \ea
 With this decomposition, the action  (\ref{action}) is transformed into
\ba
 \label{action2} S= \int d^4 x \sqrt{-g} \left [ \frac{M_P^2}{2}
R -  \frac{1}{2} \partial_\mu \rho
\partial^\mu \rho-
\frac{\rho^2}{2}  \left( \partial_\mu \theta + \e A_\mu  \right)
\left( \partial^\mu \theta + \e A^\mu  \right) - \frac{f^2(\rho)}{4}
F_{\mu \nu} F^{\mu \nu}  - V(\rho) \right]
\ea
The corresponding
Klein-Gordon equations of motion are
\ba \label{theta-Eq}
\partial_\mu\,  J^\mu =0
\\
\label{rho-Eq}
\partial_\mu \left[  \sqrt{-g} \partial^\mu \rho \right] -
\frac{ J_\mu J^\mu}{\rho^3 \sqrt{-g} }   -\sqrt{-g}\,(
V_\rho+\frac{f(\rho)f_\rho(\rho)}{2}F_{\mu \nu} F^{\mu \nu})=0 \, ,
\ea
accompanied by with the Maxwell's equation
\ba \label{Maxwell}
\partial_\mu \left(  \sqrt{-g}\, f^2(\rho)\, F^{\mu \nu} \right) = \e J^\nu \, ,
\ea
where the current $J^\nu$ is defined by
\ba J^\nu \equiv  \rho^2
\sqrt{-g} \left( \partial^\nu \theta + \e A^\nu  \right)  \, .
\ea
The conservation of $J^\mu$ from Eq. (\ref{theta-Eq}) is a
manifestation of the axial symmetry imposed on $V$. Interestingly,
Eq. (\ref{theta-Eq}) is not independent from Maxwell's equation,
where  $F^{\mu \nu}$ being anti-symmetric leads to $\partial_\mu
\partial_\nu F^{\mu \nu} = \partial_{\mu} J^{\mu}=0$.

Finally, the stress energy momentum tensor, $T_{\alpha \beta}$, for
the Einstein equation, $G_{\alpha \beta} = 8 \pi G \, T_{\alpha
\beta}$, is \ba \label{energy-momentum} T_{\alpha \beta} =
\frac{-f^2(\rho)}{4} g_{\alpha \beta} F_{\mu \nu} F^{\mu \nu}
+f^2(\rho) F_{\alpha \mu} F_\beta^{\, \mu} +  \partial_\alpha \rho
\partial_\beta \rho +  \frac{J_\alpha J_\beta}{\rho^2 | g|} -
g_{\alpha \beta} \left[ \frac{1}{2} \partial_\mu \rho
\partial^\mu \rho +  \frac{J_\mu J^\mu}{2 \rho^2 | g|}  +V
\right]\, . \ea

We are interested in the effects of a non-zero background gauge field on the evolution of system.
To fix the gauge we use the Coulomb-radiation gauge $A_0= \partial_i A^i=0$. Since our background is only time-dependent, from the constraint $J^0=0$ one concludes that $\dot \theta=0$ at the level of background. The inclusion of a non-zero background gauge field breaks the Lorentz invariance explicitly since a preferred direction is singled out in the background space-time.  We take our background gauge field to have the form  $A_{\mu}=(0,A(t),0,0)$.

As in \cite{Watanabe:2009ct} our background metric has the following form
\ba
\label{metric} ds^2 = - dt^2 + e^{2\alpha(t)}(e^{-4\sigma(t)}d x^2
+e^{2\sigma(t)}(d y^2 +d z^2)) \, .
\ea
Here $\alpha(t)$ measures the background number of e-foldings, $\dot \alpha$
represents the background isotropic Hubble expansion rate while $\dot \sigma(t)$
measure the anisotropic expansion rate. For a universe with small anisotropies, we require that
$|\dot \sigma/\dot \alpha| \ll 1$.

Assuming that the fields $\rho, A, \alpha$ and $\sigma$ are only function of
$t$ the background equations of motion are
\ba
\label{back-A-eq}
\partial_t{\left(  f^2(\rho) e^{\alpha + 4 \sigma} \dot A        \right)}& =& - \e^2 \rho^2 e^{\alpha + 4 \sigma}  A \\
\label{back-rho-eq}
\ddot\rho+3\dot \alpha\dot \rho+ V_\rho+ \left(
-f(\rho)f_\rho(\rho)\dot A^2 +\e^2 \rho A^2   \right) e^{-2\alpha+4\sigma}&=&0  \\
\label{Ein1-eq}
\frac{1}{2}\dot
\rho^2+V(\rho)+ \left(   \frac{1}{2}f^2(\rho)\dot
A^2 +\frac{\e^2\rho^2}{2}A^2 \right) e^{-2\alpha+4\sigma}
&=&
3 M_P^2 \left(   \dot \alpha^2-\dot \sigma^2 \right)  \\
\label{Ein2-eq}
V(\rho)+  \left(  \frac{1}{6}f^2(\rho)\dot
A^2+\frac{\e^2\rho^2}{3}A^2  \right)e^{-2\alpha+4\sigma}
&=& M_P^2 \left( \ddot \alpha    + 3 \dot \alpha^2 \right)  \\
\label{anisotropy-eq}
\left(\frac{1}{3}f^2(\rho)\dot A^2  -\frac{\e^2\rho^2}{3}A^2    \right) e^{-2\alpha+4\sigma}
&=& M_P^2\left( 3\dot \alpha \dot \sigma+ \ddot \sigma      \right)\, .
\ea
In the limit where $\e=0$ these equations reduce to those of \cite{Watanabe:2009ct}. One can also check that not all equations above are independent. For example,
Eq. (\ref{Ein1-eq}) can be obtained from the remaining four equations.

From Eq. (\ref{back-rho-eq}), the total energy density, ${\cal E}$, governing the dynamics  of the inflaton field is given by
\ba
\label{Veff}
{\cal E}= \frac{\dot \rho^2}{2}+V+  e^{-2\alpha+4\sigma} \left( \frac{1}{2}f^2(\rho)     \dot
A^2+\frac{\e^2\rho^2}{2} A^2 \right) \, .
\ea
Since the second and the third terms above
come from the gauge field, we refer to them respectively as the kinetic energy and potential energy associated with the gauge field.  Using Eqs.(15) and (16), the equation for acceleration of the
universe is given by
\ba
\label{accel}
\ddot\alpha +\dot\alpha^2 =
-2\dot\sigma^2-\frac{1}{3M_{P}^2}\dot
\rho^2+\frac{1}{3M_{P}^2}\left[V-\frac{1}{2}f^2(\rho)\dot
A^2e^{-2\alpha+4\sigma} \right] \, .
\ea
Interestingly, the term in effective potential proportional to $\e$ cancels out in this expression.
One also observes that for inflation to take place, corresponding to $\ddot\alpha +\dot\alpha^2 >0$, one requires that the background potential $V(\rho)$ dominates over the contribution from the gauge field kinetic energy.

In the following we are interested in configuration where inflation take places with small anisotropies such that
\ba
\label{delta}
\delta \equiv
|\frac{\dot \sigma} {\dot \alpha}| \ll 1.
\ea
Small amount of anisotropies in background inflationary dynamics may be acceptable assuming that they do not impose too much anisotropies on CMB temperature power spectrum.  In order for anisotropies to be small, the contribution of gauge field to the total energy density
should be small compared to the background potential. To parametrize this, we define the ratios
$R_1$ and $R_2$ via
\ba
\label{R12}
R_1 \equiv \frac{\dot A^2 f(\rho)^2 e^{-2 \alpha}}{2 V} \quad , \quad
R_2 \equiv \frac{ \e^2 \rho^2\, A^2  e^{-2 \alpha}}{2 V} \, .
\ea
For the contribution of the gauge field energy density into the total energy density to be small
we require $R_{1}, R_2 \ll 1$.  In this limit where the anisotropy is smaller than the slow-roll parameters (defined below), from Eq. (\ref{anisotropy-eq}) we obtains
\ba
\label{delta-R12}
\delta \simeq \frac{2}{3} (R_1 - R_2)\, .
\ea

One of our goal in this work is to see the behavior of $R_{1,2}$ during inflation. As we shall see, during early stage  inflation both $R_{1}$  and $R_{2}$ are
very small and inflation is basically driven by the background potential $V$ and one can treat
the system as isotropic inflation. As in \cite{Watanabe:2009ct}, sometime during inflation, $R_1$ rises quickly such that its contribution to the Klein-Gordon equation governing the scalar field dynamics can not be neglected. This is an attractor mechanism and as we shall see
below $R_1$ scales like the slow-roll parameters once the system is in the attractor regime.
Interestingly, the Hubble expansion rate is still predominantly given by the background potential $V$ but one should check that
the anisotropy given by Eq. (\ref{anisotropy-eq}) is under control. One new effect in our model is that sometime at the end of inflation $R_2$ becomes comparable to $R_1$ 
and the  gauge field oscillates very rapidly. Because of  the interaction term $\e^2 \rho^2 A_x^2$, the rapid oscillations of the gauge field induce rapid changes in inflaton effective mass, violating the slow-roll conditions and ending inflation abruptly. Here we would like to study these three distinct inflationary phases in some details. For convenience, we refer to these three stages of inflation as phase one, two and three, respectively. However, note hat the third inflationary stage is very short compared to other two inflationary stages.

As a measure of slow-roll parameters and 
phase change we define the dimensionless quantity $\epsilon $  and $\eta$ given by
\ba
\label{epsilon-eta}
\epsilon \equiv  -\frac{\ddot \alpha}{\dot \alpha^2} \quad , \quad 
\eta \equiv \frac{\ddot \rho}{3 \dot \alpha \dot \rho} \, .
\ea
When the slow-roll approximation holds $\epsilon , \eta \ll 1$ and  inflation ends when $\epsilon , \eta \simeq 1$.  In our anisotropic inflationary models, $\eta$ has sudden jumps which represents the onsets of phase changes. In
{\bf Fig. \ref{eta-fig}} we have plotted $\eta $ for some parameters value which clearly indicates the jumps in $\eta$.


\section{Symmetry Breaking Hilltop Inflation}
\label{symmetry}

We start with an almost isotropic configuration with negligible anisotropies such that the gauge field contributions in expansion rate Eq. (\ref{Ein1-eq}) and the Klein-Gordon equation (\ref{back-rho-eq}) are negligible, corresponding to $R_{1}, R_{2} \ll \epsilon , \eta$.  However, we would like to allow the gauge field kinetic energy to increase such that $\delta$ increases towards the  allowed observational bounds.

To be specific, we work with the symmetry breaking (Mexican hat) potential which is theoretically well motivated in a  model with an Abelian gauge field
\ba
\label{pot}
V= \frac{\lambda}{4} \left( | \phi |^2 - \frac{M^2}{\lambda}  \right)^2 
\equiv  \frac{\lambda \mu^4}{4 } \left ( \hat \rho^2 -1 \right)^2 \, .
\ea
Here $\lambda$ is a dimensionless coupling and for the later convenience we have defined the dimensionless variable $ \hat \rho \equiv \rho/\mu$ where $\mu \equiv   M/\sqrt{\lambda}$. The potential has global minima at $ \rho = \mu $ or $\hat \rho=1$.
 In this picture, we assume inflaton starts at the top of the potential
and proceeds towards the global minima. As is well  known cosmic strings are produced at the end of inflation which can have interesting observational effects.  

In \cite{Watanabe:2009ct} the authors studied the simple chaotic inflationary potential $V= m^2 \phi^2/2$ for a real scalar field. As we shall see, their conclusion about the existence of the
attractor mechanism during the second phase and the behavior of $R_1$ and $\delta$ will also hold in our case. However, the contribution of 
gauge coupling $\e$ via $R_2$ opens up new inflationary phase and the dynamics of the system at the end of inflation is quite different than what studied in \cite{Watanabe:2009ct}.

During the inflation, which happens mostly in the hilltop regions of the potential, one may approximate potential (\ref{pot}) as
\ba
V\simeq \frac{M^4}{4 \lambda} - \frac{M^2}{2}\rho^2 \, .
\ea
For this approximation to be valid, one has to satisfy $\hat \rho \ll 1$ during inflation.

Since we are interested in small anisotropies, $R_1, R_2, \delta \ll 1$,  the background expansion is given as in standard slow-roll inflationary models with
\ba
\label{Hubble}
{\dot \alpha}^2 \simeq \frac{V}{3 M_P^2}  \simeq \frac{M^4}{12 \lambda M_P^2} \, .
\ea

\begin{figure}[t]
\vspace{-0.9cm}
\includegraphics[width=9cm]{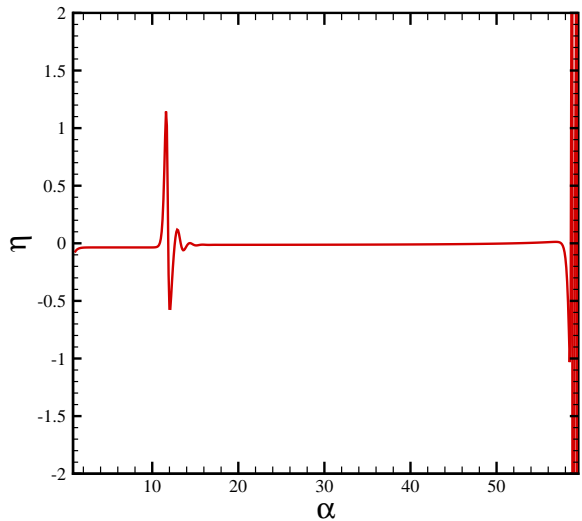}
 \caption{Here we have plotted $\eta$ defined in Eq. (\ref{epsilon-eta}), with $\lambda= 2.5 \times 10^{-13}$, $M= 3\times 10^{-6} M_P$, $p=50$, $\rho_{in} = M_p/5 $ and $\e=1$.
 The first phase change, at e-folding $\alpha\simeq 10$, happens when $R_1 $ becomes comparable to $\epsilon$. The second phase change happens very close to the end of inflation (in about one e-folding towards the end of inflation) when
 $R_2 $ also becomes comparable to $\epsilon$.}
 \vspace{0.5cm}
 \label{eta-fig}
\end{figure}

Our numerical investigations show that the phase changes happen when there are sudden changes in $\dot \rho$ in a short period. This can be seen in the plot of $\eta$ in
{\bf Fig. \ref{eta-fig} } where there are two sudden jumps. The first jump  corresponds to  the transition from phase one to phase two where the $R_1$ contribution in Eq. (\ref{back-rho-eq}) becomes important. The second jump, very close to the end of inflation,
represents the transition from the second phase to the third phase. This happens when the right hand side of Eq. (\ref{back-A-eq}) can not be neglected and it eventually affects the evolution of $\rho$ in Eq. (\ref{back-rho-eq}).  As we shall see the third period is very short
and inflation ends abruptly once the gauge field starts to oscillate during the third phase.

\subsection{Phases One and Two}

At the early stage of inflation, the contribution of gauge field in total energy density
and the scalar field equation is completely negligible corresponding to $R_{1}, R_{2} \ll \epsilon, \eta$.  Inflation proceeds as in standard slow-roll hilltop inflation and in order for the slow-roll condition to be satisfied one requires that $M_P^2 (V_\rho/V)^2$ and
$M_P^2 V_{\rho\rho}/V$ both to be much smaller than one. These in turn yields
$p_c \gg 1$ where 
\ba
\label{pc}
p_c \equiv \frac{M^2}{2 \lambda M_P^2} \, .
\ea
The Hubble expansion rate is given
by Eq. (\ref{Hubble}) while the scalar field equation in the slow-roll approximation is
\ba
\label{rho-phase1}
\rho' \simeq  4 \frac{ \lambda M_P^2}{M^2} \rho \quad   \rightarrow  \quad
\rho \simeq \rho_{in} e^{2 \alpha/p_c} \, ,
\ea
where $\rho_{in}$ represents the initial value of the inflaton field. Also here and below, prime denotes derivative with respect to $\alpha$, the number of e-foldings. We use the convention such that at the start of inflation $\alpha=0$ and the total number of e-foldings measured at the end of inflation is $\alpha= \alpha_f \simeq 60$ to solve the flatness and the horizon problem. From Eq. (\ref{rho-phase1}) the number of e-folds as a function of $\rho$  during the first phase is 
\ba 
\label{back-alpha0}
\alpha(\rho)  \simeq  \frac{p_c}{2} \ln \left(  \frac{ \rho}{ \rho_{in}}      \right)   \, .
\ea

So far we have not specified the form of $f(\rho)$, the time-dependent  gauge kinetic coupling. In order for the perturbative gauge theory to be under control we demand that the effective gauge kinetic
coupling $g_A(\alpha) = f(\rho)^{-1}$ to be small during inflation and approaches unity at the end of inflation for some yet unknown dynamical mechanism, that is $g_A(\alpha_f) =1$. This indicates that $f(\rho)$ is a decreasing function during inflation.   As mentioned before, we would like the gauge field contribution to the energy density to be subdominant but big enough to play some roles in anisotropy and scalar field equations. To determine the form of $f(\rho)$ we note that  during the first two phases, $R_2 \ll R_1$  so the terms proportional to
$\e$ in background equations (\ref{back-A-eq})-(\ref{anisotropy-eq}) can be neglected.
From Eq. ({\ref{back-A-eq}}) one obtains $\dot A  \propto f(\rho)^{-2}e^{-\alpha} $ so
$R_1 $ scales like $R_1 \propto  f(\rho)^{-2}  e^{-4 \alpha} \sim   f(\rho)^{-2} \rho^{-2 p_c}$. Therefore, for the critical coupling $f_c \equiv ( \mu/\rho)^{p_c}$ the gauge field kinetic energy remains fixed during the first two phases.
As we started with negligible $R_1$ in phase one, then it remains negligible afterwards, i.e. $R_1 \ll \epsilon$. In order to increase $R_1$ during the second phase  we consider the gauge kinetic coupling
\ba
\label{f}
f(\rho) = \left(\frac{\mu}{\rho} \right)^{p} = \hat \rho^{- p} \, ,
\ea
 with $p> p_c$
such that the gauge field kinetic energy becomes important during the second and third stages of inflation.

As explained above, during the first two phases the right hand side of Eq. ({\ref{back-A-eq}}) can be neglected and using Eq. (\ref{f})
one obtains 
\ba
\label{A-prime}
\hat A' = \left( \xi \hat \rho^2 \right)^p  e^{-\alpha} \, ,
\ea
where the dimensionless gauge field is defined via $\hat A \equiv  A/\mu$ and 
$\xi$ is a constant of integration. Note that we defined the constant of integration in this way so the subsequent analysis becomes more simplified. Physically, $\xi$ is measured by the initial value of $R_1$ at the start of inflation, $\alpha=0$.  Plugging Eq. (\ref{A-prime}) into 
Eq. (\ref{R12}) during the first two phases one obtains 
\ba
\label{R1-12}
R_1 \simeq \frac{p_c}{3} e^{-4 \alpha} \left( \xi \hat \rho \right)^{2p} \, ,
\ea
and therefore the initial value of $R_1$ is 
\ba
\label{R1in}
R_{1 \, {in}} \simeq \frac{ p_c}{3}   \left( \xi \hat \rho_{in} \right)^{2 p} \, .
\ea

Since we demand that $R_{1,2} \ll 1$, the Friedmann equation is still given by Eq. (\ref{Hubble}). Combined with Eq. (\ref{f}), the inflaton field equation in the slow-roll limit is cast into
\ba
\label{chi-eq}
(\hat \rho^2)' - \frac{4\,  \hat \rho^2}{p_c}  + \frac{2 p\,  \xi^{2p}}{3} e^{- 4 \alpha} \left(  \hat \rho^2  \right)^p = 0\, .
\ea                 

As can be seen from Eq. (\ref{R1-12}), $\left( \xi \hat \rho  \right)^{2p}$ is very small during the first phase of inflation so one can neglect the last term in Eq. (\ref{chi-eq}) and the solution is given by Eq. (\ref{rho-phase1}).  
For this to take place, we need to make sure that at the start of inflation, $\alpha=0$, the third term in  Eq. (\ref{chi-eq}) is indeed much smaller than the second term. Using the expression
of $R_{1\, in}$ given in Eq. (\ref{R1in}) this condition is transformed into
\ba
\label{cond1}
R_{1\, in} \ll \frac{2}{p} \hat \rho_{in} \, .
\ea

\begin{figure}[t]
\vspace{-0.9cm}
\includegraphics[width=8cm]{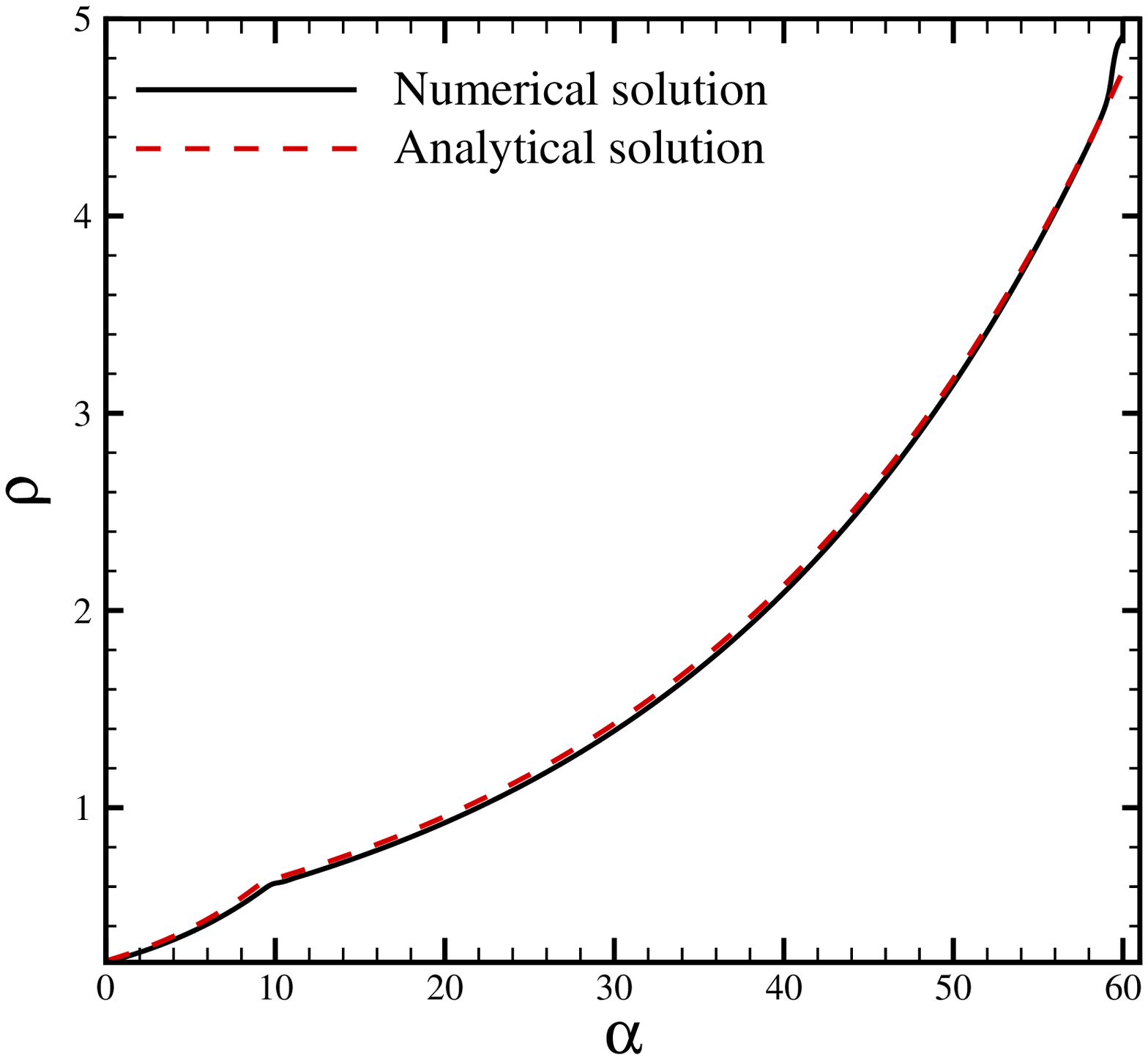} 
\includegraphics[width=8cm]{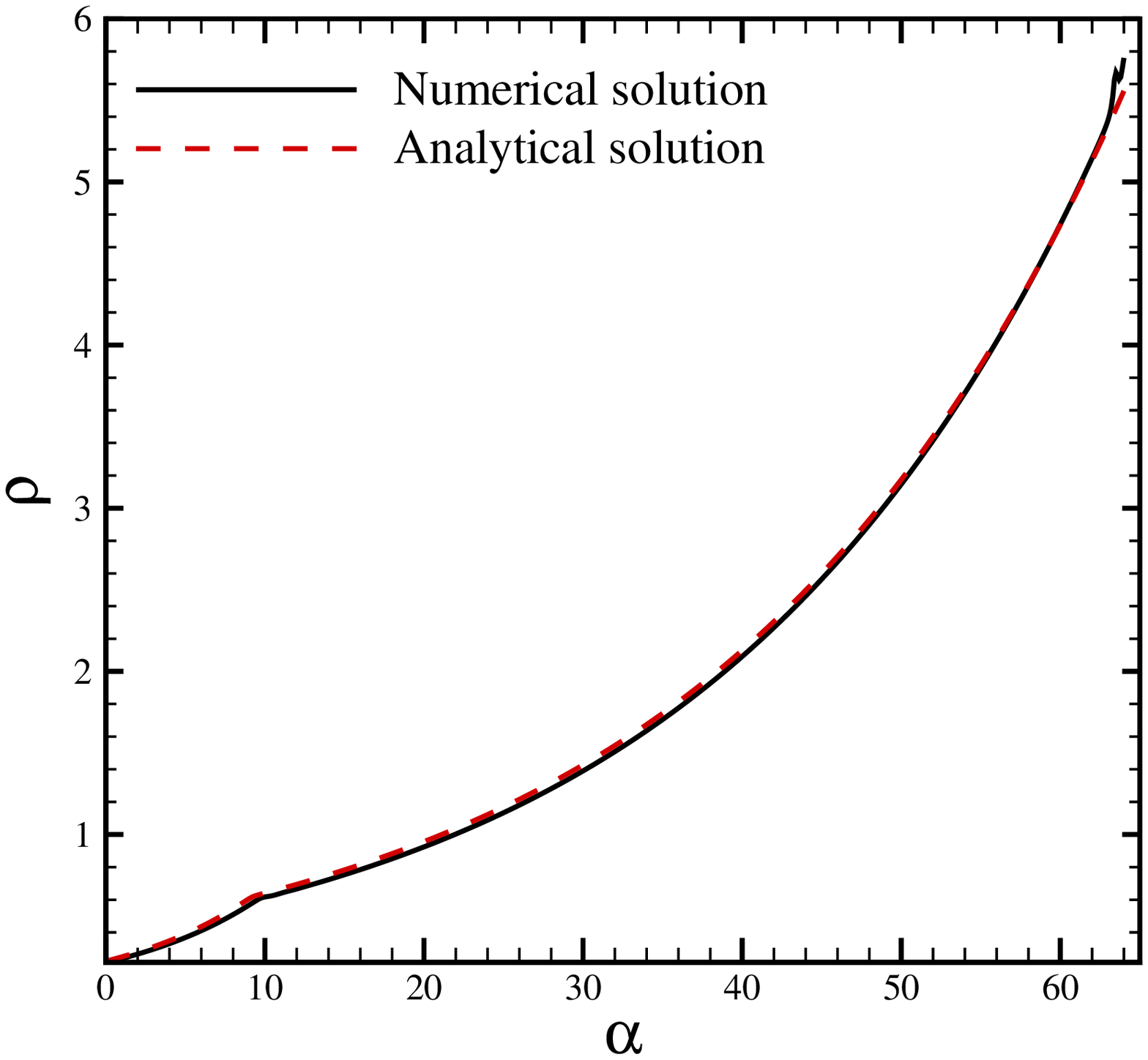}
 \caption{Here we plot our analytical solution for $\rho(\alpha)$, Eq. (\ref{hat-rho1}),
 shown by the red dashed curve, and compare it to the full numerical solution denoted by the solid black curve. The agreement between them is very good. The left figure corresponds to $\e=1$ whereas for the right figure $\e=10^{-4}$. As argued, the time of first phase change, which here is at $\alpha_1 \simeq 10$, is independent of the value of $\e$ and is well approximated by our analytical formula Eq. (\ref{alpha1}). All other parameters are as in {\bf Fig.} \ref{eta-fig}. }
 \vspace{0.5cm}
\label{r-fig-sym}
\end{figure}

As inflation proceeds and $\hat \rho$ increases the last term in Eq. (\ref{rho-phase1})
catches up with the second term and one should take the effect of this term into account. This is exactly when the gauge field contribution into the inflaton equation, Eq. (\ref{back-rho-eq}), becomes important as promised. Eq. (\ref{rho-phase1}) can be solved with the answer
\ba
\label{hat-rho1}
 \hat \rho  \simeq \frac{ \hat \rho_{in} e^{\frac{2 \alpha}{p_c}  } }{ \left[ 1 + \frac{p^2 p_c}{6 (p - p_c) } \left(  \xi \hat \rho_{in} \right)^{2p} e^{\frac{4 (p- p_c) \alpha}{p_c} }  \right]^{1/2p} }  \, .
\ea
During the first inflationary phase, the second term in the denominator is much smaller than unity and the solution to the  above equation reduces to our previous result, Eq. (\ref{rho-phase1}). The transition from the first phase to the second phase happens when the two terms in the denominator above become comparable. Defining the first phase transition to take place at $\alpha= \alpha_1$, one obtains
\ba
\label{alpha1}
\alpha_1 &\simeq& \frac{p_c}{4 (p-p_c)} \ln \left[  \frac{6 (p-p_c)}{p^2 p_c  \left( \xi \hat \rho_{in} \right)^{2p}}\right] \nonumber\\
&\simeq& \frac{ p_c}{4 (p-p_c)} \ln \left[  \frac{2 (p-p_c)}{p^2 R_{1\, in}} 
\right] \, ,
\ea
where to get the final answer Eq. (\ref{R1in}) has been used. This is an interesting result because the onset of the first phase transition is controlled by the anisotropy at the start of inflation, $R_{1\, in}$, and the parameter $p$.  As can be seen from Eq. (\ref{alpha1}), the smaller is the value of initial anisotropy  $R_{1\, in}$, the longer it takes for the system to enter 
the second inflationary regime. 
We have checked that Eq. (\ref{alpha1}) gives a good estimate of $\alpha_1$ compared to the full numerical results. 

In {\bf Fig.} \ref{r-fig-sym} we have plotted our analytical solution for $\rho(\alpha)$ compared to the full numerical analysis. As can be seen they are in very good  agreement. Since $R_2 \ll R_1$ in this period, the value of $\alpha_1$ is independent of the gauge coupling $\e$ as can be seen from the plots. The left figures correspond to $\e=1$ whereas for the right figure, $\e=10^{-4}$.

In order for the phase transition to take place during the 60 observable e-folds, we require $\alpha_1 < 60$. This in turn impose the following lower bound on $R_{1\, in}$
\ba
\label{R1-lower}
R_{1\, in} > \frac{2(p-p_c)}{p^2} \exp \left[  -\frac{240(p-p_c)}{p_c}  \right] \, .
\ea
This is reasonable, because the smaller is the initial value of anisotropy, the longer it takes for the gauge field kinetic energy to become significant to affect the dynamics of the inflaton field. For values of $p_c$ comparable to $p$  Eq. (\ref{R1-lower}) can easily be satisfied and the first phase transition takes place during the physically relevant  window of inflation.

During the second phase, $\alpha > \alpha_1$, the solution (\ref{hat-rho1}) quickly approaches to its attractor solution and
\ba
\label{hat-rho2}
 \left( \xi \hat \rho \right)^{2p} e^{-4 \alpha} \simeq \frac{6 (p-p_c)}{p^2 p_c} \, .
\ea
If we plug back this equation into the inflaton equation (\ref{chi-eq}), we find that the last term in   (\ref{chi-eq}) behaves as a constant source term with the magnitude $4 (p-p_c)/p p_c$ turned on at the time of  phase change. This explains the kink in $\rho$ behavior seen in {\bf Fig.}
\ref{r-fig-sym}.

\begin{figure}[t]
\vspace{-0.5cm}
\includegraphics[width=7.5cm]{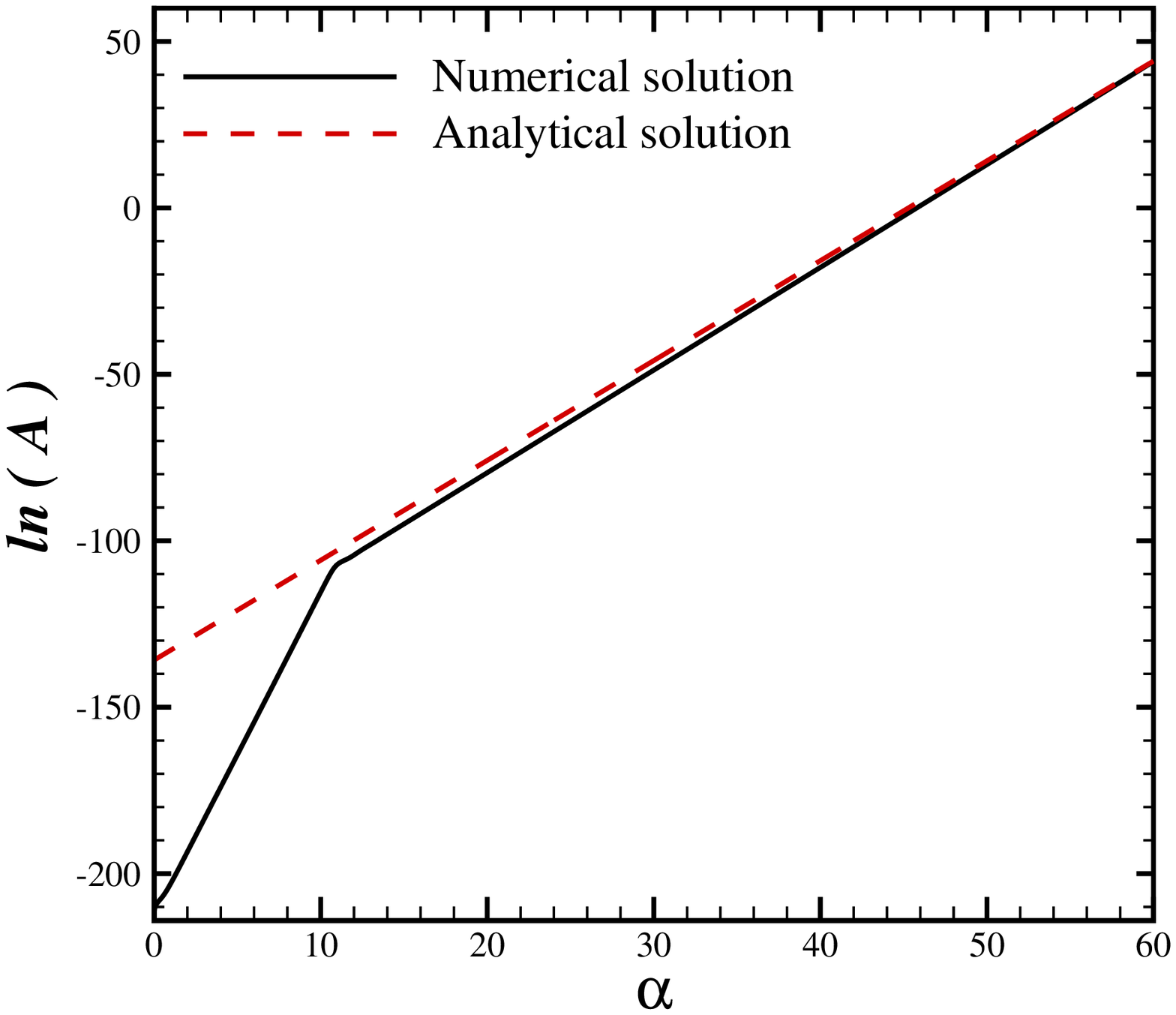} \hspace{0.5cm}
\includegraphics[width=7.5cm]{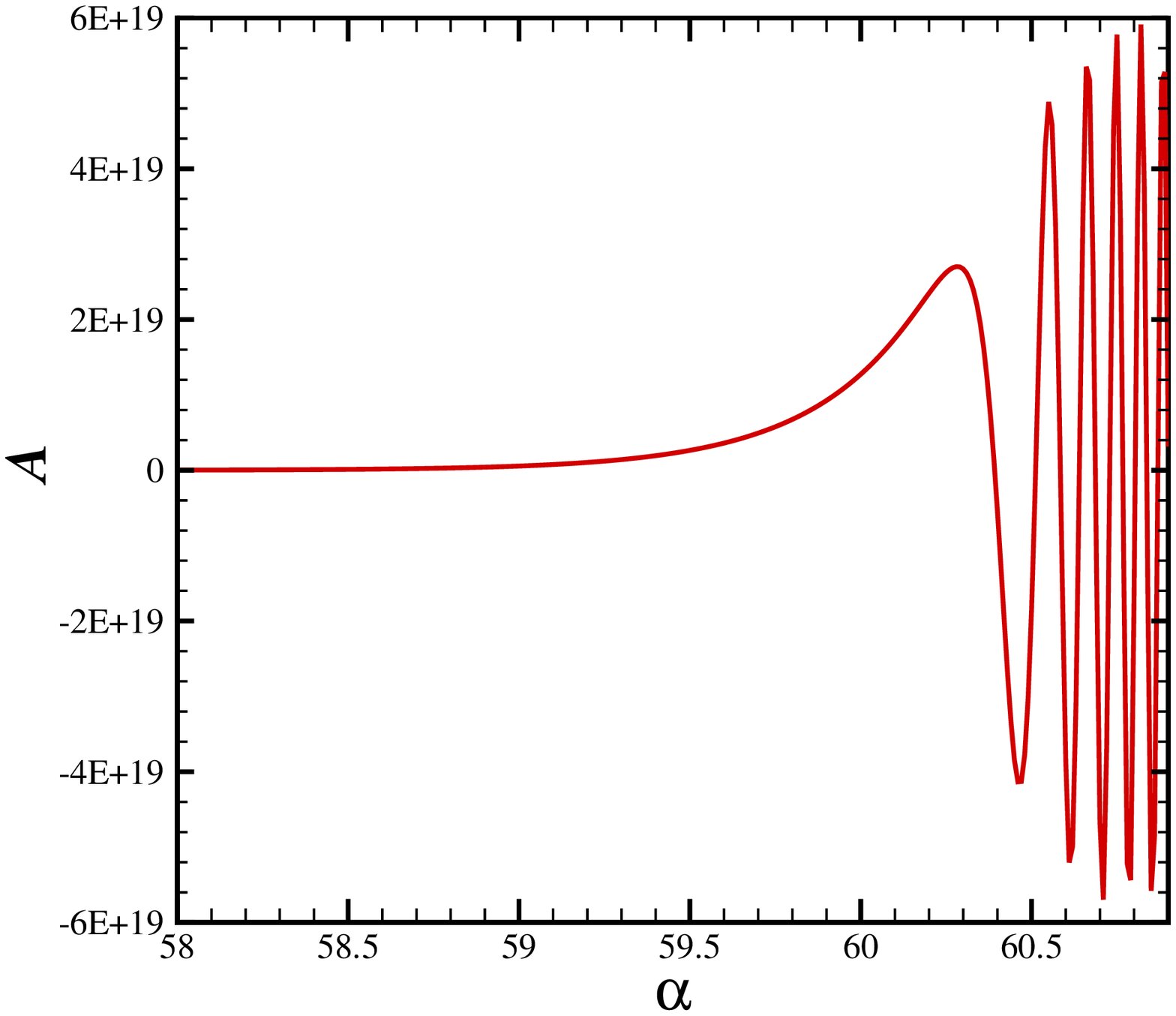}
 \caption{Here we plot  the evolution of the gauge field. The left graph represents $\ln A$ where
 the red dashed-dotted curve is our analytical solution Eq. (\ref{A3-Bes}) whereas the solid black curve is the full numerical solution. The agreement between our analytical solution Eq. (\ref{A3-Bes}) valid for the second and third phase and the full numerical result is good. Also the change of the slope of $\ln A$ form the first phase to the second phase is in good agreement with our other analytical result, Eq. (\ref{A-prime2}), valid for the first two phases. The right graph represents $A$ during the last few e-foldings. The oscillatory behavior suggested by Eq. (\ref{A3-Bes}) is clearly seen. The start of the third phase, corresponding to the first peak is well approximated by our analytical estimation Eq. (\ref{alpha2}). All parameters here are as in {\bf Fig.} \ref{eta-fig}.
 \vspace{0.5cm}
  \label{sym-A-fig} }
\end{figure}

It is also instructive to look into the gauge field evolution in this stage. Plugging Eq. (\ref{hat-rho1}) into Eq. (\ref{A-prime}) one can find an analytic expression for $\hat A'$ valid for both phase one and two with the asymptotic behavior 
\ba
\label{A-prime2}
\hat A' \simeq  \left\{
\begin{array}{c}
(\xi \hat \rho_{in}^2 )^{p} \,  \exp\left[{\frac{(4 p- p_c) \alpha}{p_c} }\right]   \quad  \quad \quad \alpha< \alpha_1
\\
 \frac{6 (p-p_c) }{p^2 p_c \xi^p} \, e^{3 \alpha} \, ~~~~~~~~~~~~~~ \quad \quad \quad \alpha> \alpha_1 
\end{array} \right. 
\ea
This indicates that the gauge field increases like $e^{3 \alpha}$ during the second phase
whereas it was increasing with a slightly higher rate, $ \exp\left[{\frac{(4 p- p_c) \alpha}{p_c} }\right]$, during the first inflationary stage. The change in the slope of the evolution of $\ln A$
clearly can be seen in {\bf Fig.} \ref{sym-A-fig}.

Finally we are in a position to find the form of $R_1$ and $\delta$. During the first inflationary stage, plugging Eqs. (\ref{A-prime2}) and (\ref{back-alpha0}) into  Eq. (\ref{R12}), yields
\ba
\label{delta-first}
\delta \simeq \frac{2}{3} R_1  \simeq \frac{2}{3} R_{1\, in} 
\exp\left[ \frac{4 (p-p_c) \alpha}{p_c} \right]  \, .
\ea
As expected, $\delta$ increases exponentially during the first inflationary stages with the initial amplitudes set by $R_{1\, in} $. During the second stage it reaches its  attractor value. Plugging Eq.  (\ref{hat-rho2}) into Eq. (\ref{R12}), during the attractor regime we have
\ba
\label{R1-hilltop}
 \delta \simeq \frac{2R_1}{3} \simeq \frac{4(p-p_c)}{3p^2}.
\ea
This attractor value is fairly independent of  the initial conditions.  

In chaotic model studies in \cite{Watanabe:2009ct}  it was shown that $R_1$ follows the slow-roll parameters $R_1 \sim \epsilon$. Here we show this conclusion also holds for our case. To see this, note that in the slow-roll limit
$
\epsilon \simeq 2 R_1 + \frac{3\dot \rho^2}{2 V}.
$
Using Eq. (\ref{hat-rho2}) in Eq. (\ref{chi-eq}) one can approximately find that   
$\epsilon \simeq 2 R_1 + 4 \hat \rho^2 p_c/p^2$. Since  $\hat \rho <1$ during the second phase 
one concludes that 
\ba
\label{epsilon-R1}
\epsilon \gtrsim 2 R_1 \, .
\ea

\begin{figure}[t]
\vspace{-0.5cm}
\includegraphics[width=8cm]{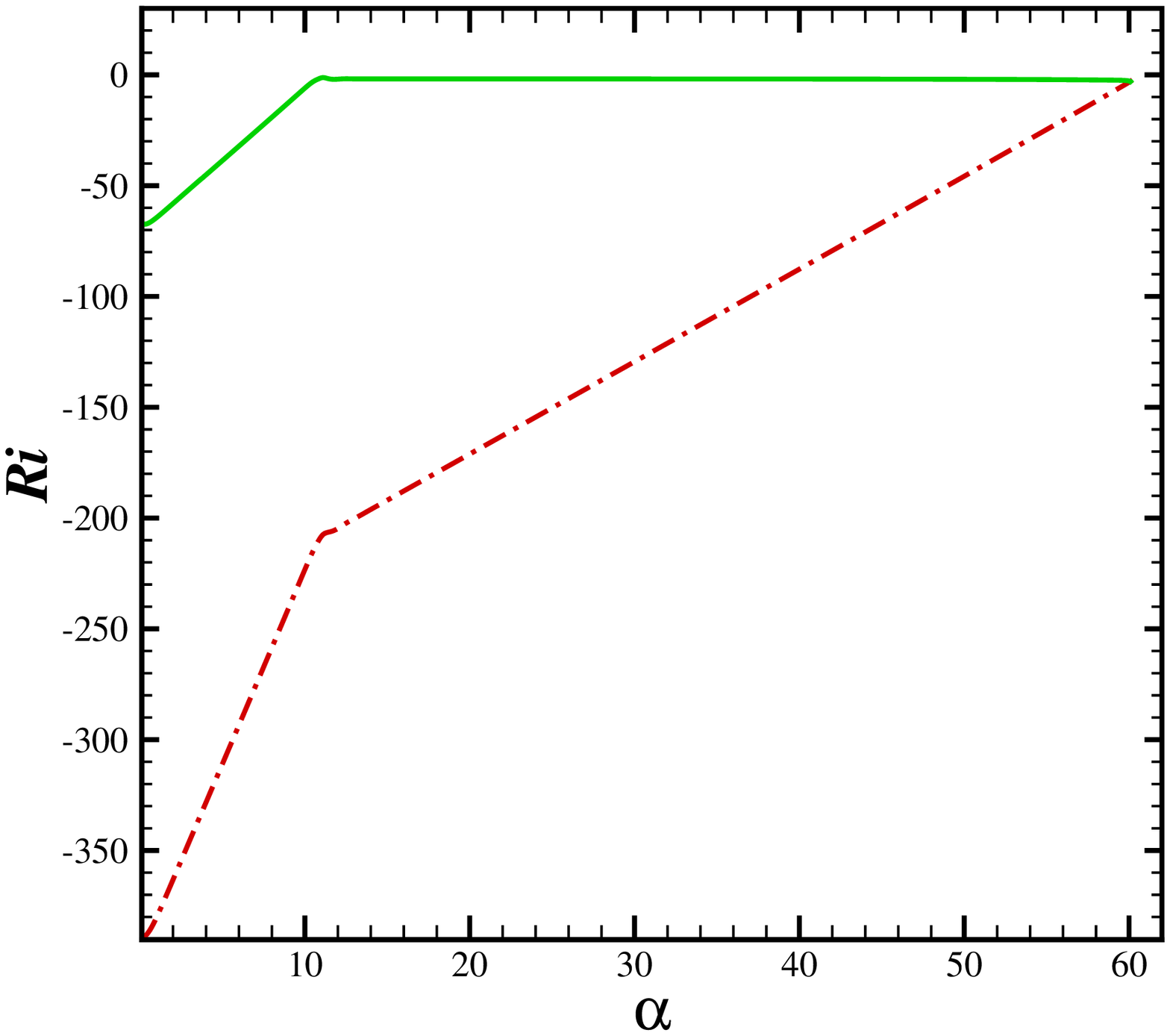} \hspace{0cm}
\includegraphics[width=8cm]{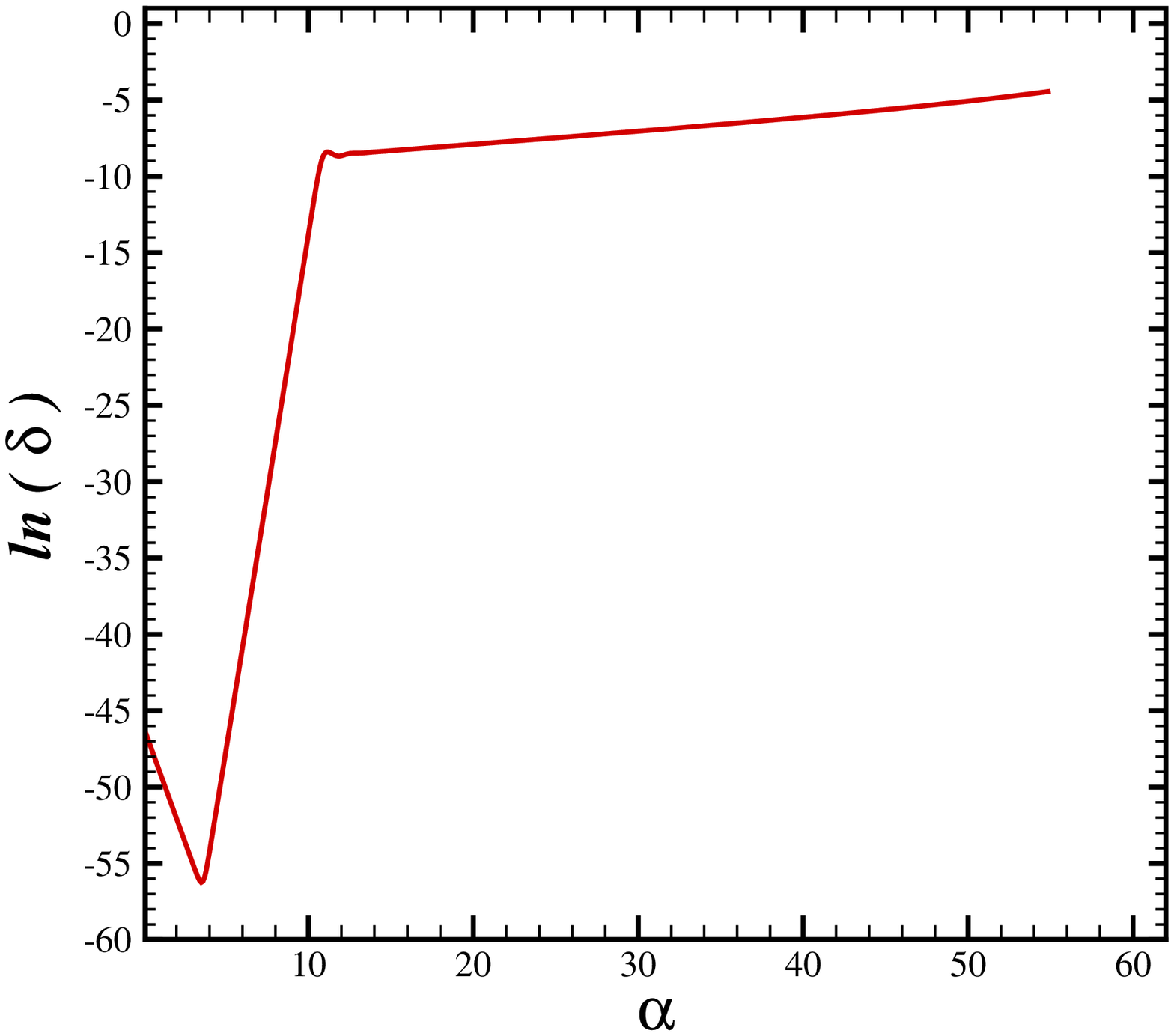} \hspace{1cm}
 \caption{In left figure we plot  $\ln (R_1/\epsilon) $ ( upper solid green curve) and $\ln( R_2/\epsilon)$ (lower dashed-dotted red curve). The phase change takes place at $\alpha_1 \simeq 10$ followed by the attractor regime denoted by the almost horizontal line where $ R_1 \propto \epsilon $.
 As explained in the text, $R_2$ is very small compared to $R_1$ until the very end of inflation when they become comparable and inflation ends shortly after that. Right: $\ln \delta$ is presented. The attractor behavior during the second inflationary stage is clear.
 All parameters here are as in {\bf Fig.} \ref{eta-fig}. }
 \vspace{0.5cm}
\label{Ri-delta-fig}
\end{figure}

In {\bf Fig.} \ref{Ri-delta-fig} we have plotted the ratio $R_1/\epsilon$ and the anisotropy $\delta$. As can be seen, our analytical formulae Eqs. (\ref{epsilon-R1})  and (\ref{R1-hilltop}) are in good agreement with the full numerical results. As explained before, the solution during the second phase quickly reaches the attractor regime where $R_1\simeq \epsilon \simeq \delta$ and the fraction of gauge field energy density to the total energy density  in the Friedmann equation is at the level of slow-roll parameter.  The attractor phase can clearly be seen from the behavior of $R_1$ and $\delta$ in {\bf Fig.} \ref{Ri-delta-fig}.

As the gauge field $A$ increases exponentially the effective potential for the inflaton field increases as $\e^2 A^2 \rho^2 e^{-2 \alpha} \propto \e^2  \rho^2 e^{4 \alpha}$ and  the slow-roll condition quickly terminates at the final stage of inflation. This is the third stage where $R_2 $ becomes comparable to $R_1$. Below we study this phase in some details.

\subsection{Final stage of inflation}

Now we consider the final stage of inflation when the right hand side of Eq. (\ref{back-A-eq}) can not be neglected. Using Eq. (\ref{hat-rho2}) in Eq. (\ref{back-A-eq}), the equation of motion for the gauge field can be approximated to
\ba
\label{A-eq1}
\hat A''  - 3  \hat A' +  \beta  e^{4 \alpha} \hat A=0
\ea
where the dimensionless parameter $ \beta$ is defined via
\ba
\label{beta}
 \beta \equiv \frac{36 \e^2 \, (p-p_c)}{\lambda \, p^2 p_c^2 \, \xi^{2p}} \, .
\ea
The solution to this equation is in the form of Bessel functions 
\ba
\label{A3-Bes}
\hat A = e^{\frac{3 \alpha}{2}} \left[ a_1 J_{3/4} \left(  \frac{ \sqrt { \beta}}{2} e^{2 \alpha}  \right) + a_2 Y_{3/4} \left(  \frac{ \sqrt { \beta}}{2} e^{2 \alpha}   \right) \right] \, ,
\ea
where $a_1$ and $a_2$ are the constants of integration. 

From the form of Eq. (\ref{A-eq1}) it is seen that the third inflationary phase starts when the last term in Eq. (\ref{A-eq1}) is comparable to the second term. This means that 
$\sqrt \beta e^{2\alpha_2} \simeq 1$ where $\alpha=\alpha_2$ is the start of the third inflationary stage. This gives 
\ba
\label{alpha2}
\alpha_2 \simeq  \frac{1}{4} \ln \left[ \frac{\lambda \, p^2 p_c^2\,  \xi^{2p}}{36\,  \e^2 \, (p-p_c)} \right] \, .
\ea
We have checked numerically that this expression gives a very good estimate for $\alpha_2$, the onset of transition from the second inflationary stage to the third inflationary stage. Shortly after $\alpha > \alpha_2$, the argument of the Bessel function  exponentially increases and  the gauge field starts to oscillate. This in turn triggers 
a sharp increase in the slow-roll parameters $\epsilon$ and $\eta$ and inflation ends abruptly. 
This can be seen  in the plot of $\eta$ shown in {\bf Fig.} \ref{eta-fig}.

For the consistency of our setup we require that $\alpha_1 < \alpha_2$, i.e. the third inflationary stage takes place after the second inflationary stages. Comparing Eqs. (\ref{alpha2}) and (\ref{alpha1}), and assuming $(p-p_c)/p_c \sim 1$, one requires
\ba
\e^2 \ll \lambda p^4 \xi^{2p} R_{1\, in} \, .
\ea
Because $\hat \rho_{in} \ll 1$ and $p\gg 1$, this condition can easily be met.

The condition $\sqrt \beta e^{2\alpha_2} \simeq 1$  indicates that during the final stage of inflation the arguments of the Bessel functions in Eq. (\ref{A3-Bes}) are bigger than unity while during the first two inflationary stage the arguments of the Bessel functions are small. Using the small argument limit of the Bessel function,  $J_{3/4} (x)  \sim  x^{3/4}$ and $Y_{3/4}(x) \sim x^{-3/4}$ for $x\ll 1$ one concludes that the term containing $Y_{3/4}$ decays quickly as inflation proceeds and the term containing $J_{3/4}$ survives in Eq. (\ref{A3-Bes}). 
More specifically, $J_{3/4} \left( \sqrt \beta\,  e^{2\alpha}/2 \right) \simeq 
(\sqrt \beta/2)^{3/4}  e^{3 \alpha/2}$ and comparing this with Eq. (\ref{A-prime2}) during the second inflationary stage  one can fix  the coefficient $a_1$ to obtain
\ba
\label{A3-Bes}
\hat A \simeq  \frac{2^{5/4} (p-p_c)}{ p^2 p_c \, \xi^p \, \beta^{3/8}} \, 
e^{3 \alpha/2}  J_{3/4} \left(  \frac{ \sqrt { \beta}}{2} e^{2 \alpha}  \right) \, .
\ea
Note that this expression works for the second and the third inflationary stages whereas the formula Eq. (\ref{A-prime2}) works for the first two inflationary stages. We have checked that 
Eq. (\ref{A3-Bes}) is in good qualitative agreement with the full numerical analysis. In {\bf Fig}
\ref{sym-A-fig} we have compared Eq. (\ref{A3-Bes}) with the full numerical result and the agreement between them is good. Also in {\bf Fig}
\ref{sym-A-fig} we have plotted the behavior of $A$ for the last few e-foldings. The start of the third phase is when the argument of the Bessel function in Eq. (\ref{A3-Bes}) becomes at the order of unity given by Eq. (\ref{alpha2}). This corresponds to the first peak in the plot of $A$
in the right figure of {\bf Fig} \ref{sym-A-fig}.

It is also instructive to compare $R_2$ with $R_1$ during the final stage of inflation. 
As explained before, $R_1$ and $R_2$ measure respectively the gauge field kinetic energy and potential energy compared to the background inflationary potential. Physically, we expect that during the final stage of inflation $R_2$ rises quickly and becomes comparable to $R_1$ and $\epsilon$. 
In {\bf Fig} \ref{Ri-delta-fig} we have compared $R_2$ with $R_1$ and $\delta$. Initially 
$R_2$ is very small, but during the final stage of inflation $R_2$ rises quickly and becomes comparable to $R_1$. Physically this means that the inflaton mass receives a time-dependent contribution of the form $\e^2 A^2 \rho^2 e^{-2 \alpha}$ and the slow-roll conditions are violated soon after the gauge field starts to oscillate.
This  conclusion is supported in both {\bf Fig} \ref{sym-A-fig} and {\bf Fig} \ref{eta-fig}.

Finally it is also instructive to look into the behavior of the inflaton field as a function of the the strength of the gauge coupling $\e$. As argued before, during the first two inflationary stages 
$\e$ does not play important roles and the evolution of the inflation proceeds as in $\e=0$. In particular, the position of the first kink, $\alpha_1$, is quite insensitive to the value of $\e$.
On the other hand, $\e$ controls the end of the second inflationary  phase, $\alpha_2$, but only logarithmically. In the right plot of {\bf Fig.} \ref{r-fig-sym} we have changed $\e$ by four orders of magnitudes. Correspondingly, $\alpha_2$, and the total number of e-foldings 
changed by 4. This is consistent with our analytical formula Eq. ({\ref{alpha2}})
which for $\e \rightarrow 10^{-4} \e$  predicts an increase of e-foldings of 
$( \ln 10^{8})/4 \simeq 4.6 $.

\section{Charged Hybrid Inflation}
\label{hybrid}

In the previous example the inflaton field was a complex field and was responsible for the symmetry breaking. Now we consider the case where the inflaton field is real, while the symmetry breaking 
is controlled by another complex scalar field, the waterfall field. The action is
\ba
\label{action3} S=\int d^4 x  \sqrt{-g} \left [ \frac{M_P^2}{2} R - \frac{1}{2} \partial_\mu \phi
\,  ^\mu\phi- \frac{1}{2} D_\mu \psi
\,  D^\mu\bar\psi - \frac{f^{2}(\phi)}{4} F_{\mu \nu} F^{\mu
\nu}- V(\phi, \psi, \bar \psi) \right]  \, .
\ea
As explained above, $\phi$ is the inflaton field while $\psi$ is the complex waterfall field. 

The potential is as in standard hybrid inflation \cite{Linde:1993cn}
\ba
\label{pot3} V(\phi, \psi, \bar \psi)=\frac{\lambda}{4 }  \left(  |\psi|^2 - \frac{M^2}{\lambda} \right)^2 + \frac{g^2}{2} \phi^2 |\psi|^2 + \frac{m^2}{2} \phi^2 \, .
\ea   
We are interested in the configuration where the potential is axially symmetric and $V(\psi, \bar \psi , \phi)= V(\chi, \phi)$ where $ \psi(x) = \chi(x) \,  e^{i \theta(x)}$. 
Following the same metric ansatz as in Eq. (\ref{metric}) and taking $A_{\mu}=(0,A(t),0,0)$ the equations of motion are
\ba
\label{back-A-eq3}
\partial_t{\left(  f^2(\phi) e^{\alpha + 4 \sigma} \dot A \right)}& =& - \e^2 \chi^2 e^{\alpha + 4 \sigma}  A \\
\label{back-phi-eq3}
\ddot\phi+3\dot \alpha\dot \phi+ \phi(m^2+g^2\chi^2) -f(\phi)f_\phi(\phi)\dot A^2 e^{-2\alpha+4\sigma}&=&0  \\
\label{back-chi-eq3}
\ddot\chi+3\dot \alpha\dot \chi+ \left(\frac{\lambda}{4}(\chi^2-\frac{M^2}{\lambda})+g^2\phi^2 \right)\chi
+e^2\chi A^2 e^{-2\alpha+4\sigma}&=&0  \\
\label{Ein1-eq3}
\frac{1}{2}\dot
\phi^2+\frac{1}{2}\dot \chi^2+V(\phi,\chi)+ \left( \frac{1}{2}f^2(\phi)\dot
A^2 +\frac{\e^2\chi^2}{2}A^2 \right) e^{-2\alpha+4\sigma}
&=&
3 M_P^2 \left(\dot \alpha^2-\dot \sigma^2 \right)  \\
\label{Ein2-eq3}
V(\phi,\chi)+  \left(  \frac{1}{6}f^2(\phi)\dot
A^2+\frac{\e^2\chi^2}{3}A^2  \right)e^{-2\alpha+4\sigma}
&=& M_P^2 \left( \ddot \alpha    + 3 \dot \alpha^2 \right)  \\
\label{anisotropy-eq3}
\left(\frac{1}{3}f^2(\phi)\dot A^2  -\frac{\e^2\chi^2}{3}A^2 \right) e^{-2\alpha+4\sigma}
&=& M_P^2\left( 3\dot \alpha \dot \sigma+ \ddot \sigma\right)\, .
\ea 

From Eq. (\ref{Ein1-eq3}), the total energy density determining  the expansion rate of the universe is given by
\ba
\label{energy2}
{\cal E}=V(\phi,\chi) + e^{-2\alpha+4\sigma} \left( \frac{1}{2}f^2(\phi)     \dot
A^2+\frac{\e^2\chi^2}{2} A^2 \right) \, .
\ea

The interesting new effect is that the gauge coupling $\e$ induces a new time-dependent mass term for the waterfall field in the form $\e^2 e^{-2 \alpha} A^2 \,  \chi^2$. This can be seen both
in total energy density and also in equation governing the dynamics of the waterfall field, Eq. (\ref{back-chi-eq3}).  As in standard hybrid inflation we work in the vacuum dominated regime where the waterfall field is very heavy during inflation so $\chi$ quickly settles down to its instantaneous minimum $\chi=0$ during inflation. In standard hybrid inflation  models, inflation ends when  inflaton field reaches a critical value,
$\phi=\phi_c \equiv \frac{M}{g}$, where the waterfall field becomes tachyonic and rolls down very quickly to its global minimum $\psi=\mu\equiv M/\sqrt{\lambda}, \phi=0$ ending inflation very efficiently. In our model to find the moment when the waterfall field becomes tachyonic, let us calculate the $\chi$ field  effective mass
\ba
\label{chi-mass}
\frac{\partial^2 V}{\partial \chi^2}\large|_{\chi=0}  = g^2 (\phi^2 - \phi_c^2) + \e^2 e^{- 2 \alpha} A^2 \, .
\ea
 In the absence of the gauge field, the onset of waterfall field instability is when $\phi= \phi_c$. However, in the presence of the gauge field the time when the  tachyonic instability is triggered 
is modified. Indeed, if either of $\e$  or the background gauge field $A$ are very large, then the onset of tachyonic instability can be significantly altered and inflation will end before $\phi$ reaches $\phi_c$. This can have profound effects on the dynamics of waterfall phase transition and symmetry breaking \cite{Felder:2000hj, Dufaux:2010cf}. Furthermore, one needs to revisit the question of tachyonic preheating in this case.

 The condition of waterfall phase transition, Eq. (\ref{chi-mass}), can be rewritten as
\ba
\label{transition}
\hat \phi^2 + \frac{\e^2}{g^2} \hat A^2 e^{-2\alpha} -1=0
\ea
where we have defined the dimensionless fields $\hat \phi \equiv \phi/\phi_c$ and 
$\hat A \equiv A/\phi_c$. In this notation, in the absence of gauge field the waterfall phase transition happens at $\hat \phi=1$. If the gauge field is expected to play important roles in determining the dynamic of waterfall phase transition then one requires the second term in Eq. (\ref{transition}) to become at the order of unity at the time of transition. Below we will study under what conditions on model parameters this condition can be met.

As mentioned above, we assume the waterfall  field is very heavy during inflation and the potential driving inflation is 
\ba
V\simeq \frac{M^4}{4\lambda}+\frac{1}{2}m^2\phi^2  \, .
\ea
In order for the inflaton field to be light during inflation so the slow-roll conditions are met we need $p_c\gg1$ where now $p_c$ is defined via 
\ba
p_c \equiv \frac{M^4}{2 \lambda m^2 M_P^2} \, .
\ea
Furthermore, the assumption that the waterfall field is very heavy during inflation requires
$\lambda M_P^2 /M^2 \gg1$. Finally, the condition of vacuum domination during inflation 
is met if $\lambda/g^2 \ll M^2/m^2$.

As in our previous symmetry breaking example we assume the anisotropies are negligible corresponding to $\delta \lesssim \epsilon$ throughout inflation 
so the background expansion is still given by Eq. (\ref{Hubble}). 
As in symmetry breaking example, inflation starts with the isotropic limit where $R_1 \ll \epsilon$ at the early stage of inflation. As inflation proceeds, $R_1$ rises quickly and we enter the second phase of inflation where the gauge field dynamics affect the evolution of the inflaton field in Eq. (\ref{back-phi-eq3}). However, unlike the previous example, the final stage of inflation will be very different where now inflation ends violently once the waterfall field becomes tachyonic.

During the first phase of inflation, the inflaton dynamics is
\ba
\label{phase1-hybrid}
\phi' + \frac{2 \phi}{p_c} =0 \rightarrow  \phi \simeq \phi_{in} e^{-2 \alpha/p_c}  \, ,
\ea
where $\phi_{in}$ is the initial value of the inflaton field.  The number of e-foldings is 
\ba
\alpha = -\frac{p_c}{2} \ln \left( \frac{\phi}{\phi_{in}} \right)  \, .
\ea

Now we need to determine the form of the gauge kinetic coupling, $f(\phi)$, such that 
$R_1$ rises quickly during the second phase of inflation. Since during inflation $\chi=0$, then 
the gauge field equation (\ref{back-A-eq3}) can easily be solved with $A' \sim e^{-\alpha} f(\phi)^{-2}$. Consequently, $R_1$ scales like $R_1 \sim f(\phi)^{-2} e^{-4 \alpha} \sim 
f(\phi)^{-2} \phi^{2 p_c}$. Therefore, for the critical coupling $f_c \sim \phi^{p_c}$,  $R_1$
remains fixed and the fraction of the gauge field kinetic energy to the background energy density remains fixed. As in previous example, we allow for the following coupling
\ba
\label{f2}
f(\phi) =\left(  \frac{\phi}{\phi_c}\right)^p = \hat \phi^{p} \, ,
\ea
with $p> p_c$ so the energy density of the gauge field increases as inflation proceeds. As in the symmetry breaking case, we expect to enter the attractor regime where $R_1 \simeq \epsilon$ till inflation ends via tachyonic instability. With gauge kinetic coupling given by Eq. (\ref{f2}), the gauge field equation can be solved easily and
\ba
\label{A-prime-hybrid}
\hat A' = e^{-\alpha} \left( \xi \hat \phi^2 \right)^{-p} \, ,
\ea
where $\xi$ is a constant of integration.  Plugging  the gauge field solution into the inflaton field equation yields
\ba
(\hat \phi^2)' + \frac{4 \hat \phi^2}{p_c}  - \frac{2p \,  \xi^{-2p }}{3} ( \hat \phi^2)^{-p} e^{-4\alpha} =0 \, .
\ea
In comparison to symmetry breaking analysis, it is interesting to note that 
one can reproduce the previous results with the replacements $p \rightarrow -p$ and $p_c \rightarrow -p_c $.  Like in symmetry breaking example, this equation can be solved easily with the solution during the first stage of inflation given by Eq. (\ref{phase1-hybrid}) while during the second stage of inflation, for $\alpha > \alpha_1$, one obtains the attractor solution
\ba
\label{hat-phi-eq}
\left( \xi \hat \phi \right)^{2p} e^{4 \alpha} \simeq \frac{p^2 p_c}{6 (p-p_c)} \, .
\ea
Furthermore,  plugging Eq. (\ref{hat-phi-eq}) into gauge field equation (\ref{A-prime-hybrid}) yields
\ba
\label{A-hybrid}
\hat A \simeq \frac{2 (p-p_c) \xi^{p}}{p^2 p_c} e^{3\alpha} \, .
\ea
As expected, the gauge field increases exponentially.

We can also calculate the level of anisotropy in this scenario. For the first inflationary stage, $\delta$ is given as in Eq. (\ref{delta-first}) whereas during the second stage it reaches the attractor value
\ba
\label{R1-hybrid}
\delta \simeq \frac{2}{3} R_1 \simeq 
 \frac{4(p-p_c)}{3 p^2} \frac{\lambda m^2}{g^2 M^2} \, .
\ea
As expected, $R_1$ reaches the scaling solution during the second inflationary stage.  Like in symmetry breaking case one obtains $\epsilon \gtrsim 2 R_1$.

Now we have all the tools to answer our original question that under what conditions the gauge field can play a role in triggering the water field phase transition. As explained below Eq. (\ref{transition}), this can happen when the combination $(\e^2/g^2) \hat A^2 e^{-2\alpha_f}$
is comparable to unity at the end of inflation when $\alpha= \alpha_f \simeq 60$ and $\hat \phi=1$. Using Eqs. (\ref{hat-phi-eq}) and (\ref{A-hybrid}),  and noting that by definition 
$\hat \phi=1$ when $\alpha= \alpha_f$, one obtains the interesting results 
\ba
\frac{\e^2}{g^2} \hat A^2 e^{-2\alpha}\large|_{\alpha_f} \simeq  \frac{2 \e^2 (p-p_c)}{3g^2 p^2 p_c} \sim \frac{\e^2}{p^2 g^2} \, .
\ea
This indicates that for for $\e \ll p \, g$ the gauge field does not play important role in triggering the waterfall field tachyonic instability and inflation ends as in standard hybrid inflation. However, for $\e \gg p \, g$, then the gauge field shuts off inflation before $\phi =\phi_c$
and the dynamics of the waterfall phase transition, symmetry breaking and tachyonic preheating would be drastically different than what happens in standard hybrid inflation. Because of the inhomogeneous end of inflation  large non-Gaussianities can be produced at the end of inflation in the light of \cite{Yokoyama:2008xw}. We would like to come back to this question in a future publication.

\section{Chaotic Inflation}
\label{chaotic}

As our final example, here we briefly study the case of chaotic inflation. Many of our previous results also apply here. For this purpose, we will be brief here only emphasizing our main results  and compare them to the results of \cite{Watanabe:2009ct} where they assumed $\e=0$.  
To be specific, we concentrate on the quadratic potential 
\ba
V= \frac{m^2}{2} |\phi|^2 \, .
\ea 
The equation of motions are the same as in symmetry breaking example.  During the first inflationary stage, the relevant equations are
\ba
\label{back-chaotic}
\dot \alpha^2 \simeq \frac{m^2 \rho^2}{6 M_p^2} \quad , \quad
3 \dot \alpha \dot \rho + m^2 \rho \simeq 0 \, ,
\ea 
where $\rho \equiv | \phi|$. Note that in our slow-roll limit the Friedmann equation above also holds throughout the inflationary period. These equations can easily be solved to give
\ba
\label{rho-chaotic1}
\rho^2 \simeq \rho_{in}^2 - 4 M_P^2 \alpha \, .
\ea

Now we determine the form of the desired gauge kinetic coupling. As in previous examples 
we start with an almost isotropic configuration with negligible anisotropies such that the gauge field contributions in expansion rate Eq. (\ref{Ein1-eq}) and the Klein-Gordon equation (\ref{back-rho-eq}) are negligible. However, we would like to allow the gauge field kinetic energy to increase such that $\delta$ is within the observational bounds. As  can also be seen from the solution of Eq. (\ref{back-A-eq}), in order for the gauge field energy density to remain constant during the first inflationary stage  the gauge kinetic coupling should have the critical form $f_c \sim e^{2 \alpha}$. For the power law inflationary potentials with $V \propto \rho^n$ the critical gauge kinetic coupling  is given by $f_c \sim e^{ \rho^2 /n M_P^2}$. However, as in \cite{Watanabe:2009ct},  for the energy density of the gauge field to increases during the course of inflation we consider $f = e^{c \rho^2/n M_P^2}$ with $c>1$. For our specific example with $n=2$ we consider the gauge coupling 
\ba 
f = e^{c \rho^2/2M_P^2} \, .
\ea

During the first two inflationary stage one can neglect the right hand side of Eq. (\ref{back-A-eq}) and the gauge field evolution is given by
\ba
\label{Adot-chaotic}
\dot A = p_A \exp\left[  - \alpha -  \frac{c\rho^2}{M_p^2} \right] \, ,
\ea
where $p_A$ is a constant of integration defined in \cite{Watanabe:2009ct}. Plugging this into the inflaton equation results in
\ba
\label{rho-chaotic-eq2}
( \rho^2)' + 4 M_P^2  - \frac{4 c \, p_A^2}{m^2}  \exp\left[  -4 \alpha -  \frac{c\rho^2}{M_p^2} \right] = 0\, .
\ea 
 
\begin{figure}[t]
\includegraphics[width=7.5cm]{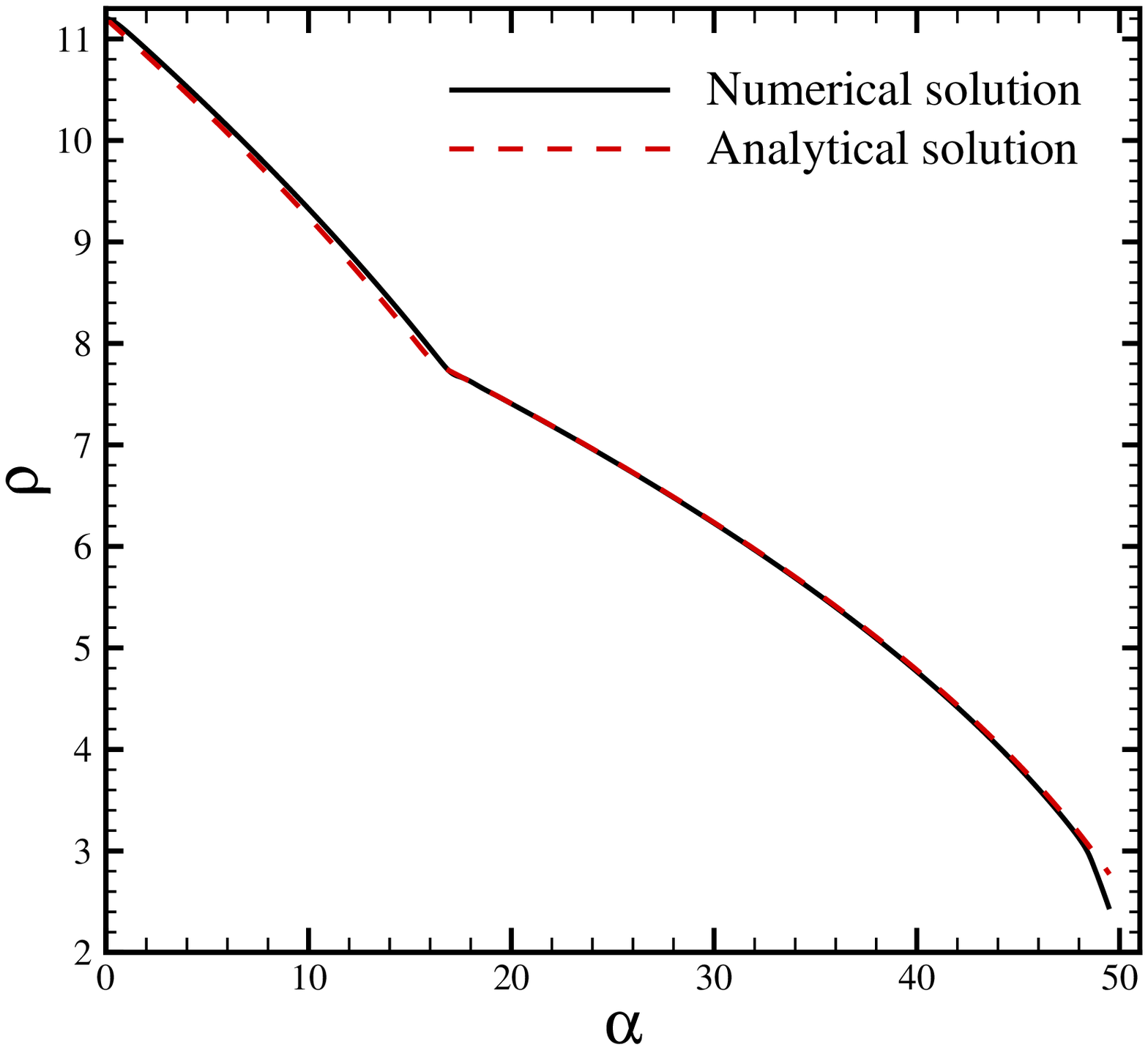} \hspace{1cm}
\includegraphics[width=7.5cm]{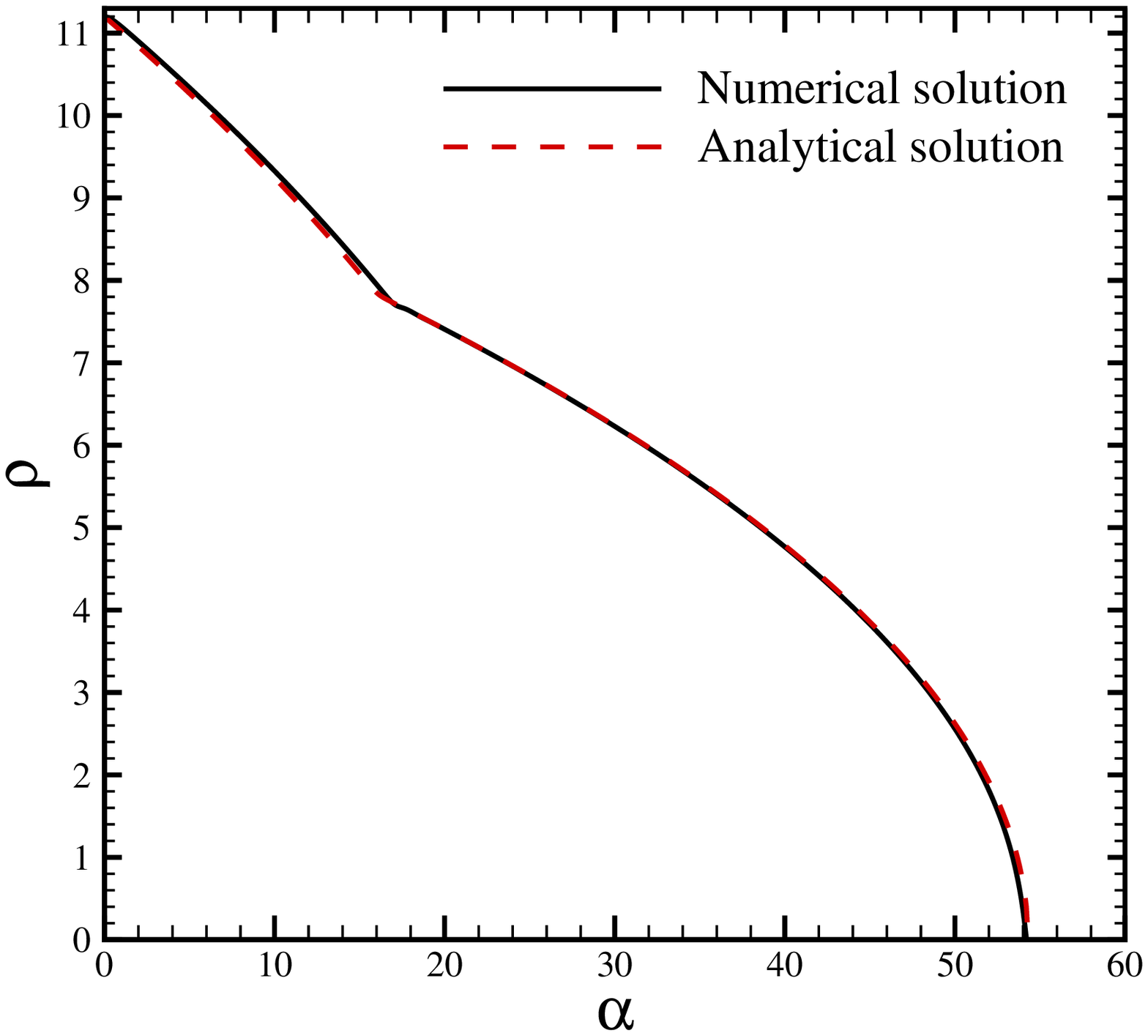}
 \caption{Here we plot the evolution of $\rho(\alpha)$ in chaotic inflation. The dashed-dotted red curve is the analytical solution whereas the solid black curve is the full numerical result.
 As can be seen, the agreement between them is very good. The left figure corresponds to $\e=0.1$ whereas for the right figure $\e=0$.  As explained below, the position of the first kink is independent of the value of $\e$ and is well approximated by Eq. (\ref{alpha1-chaotic}). 
 However, the total number of e-foldings  depends logarithmically  on $\e$. Other parameters are $m=10^{-6} M_P$, $\rho_{in} = 11.2 M_P$ and $c=2.5$. }
 \vspace{0.5cm}
\label{r-fig-chaotic}
\end{figure}
           
During the first inflationary stage, when the third term in Eq. (\ref{rho-chaotic-eq2}) is negligible compared to the second term,  the solution is given  by Eq. (\ref{rho-chaotic1}). The phase change happens when the last term above becomes comparable to the second term. Because of the attractor mechanism the solution rapidly converges to
\ba
\label{rho-chaotic2}
\exp\left(  -4 \alpha -  \frac{c\rho^2}{M_p^2} \right) \simeq \frac{m^2 M_P^2 (c-1)}{c^2 p_A^2} \, .
\ea
Note that this solution is analogous of Eq. (\ref{hat-rho2}) for the symmetry breaking example.
Now plugging Eq. (\ref{rho-chaotic2}) into Eq. (\ref{rho-chaotic-eq2}) yields
\ba
3 \dot \alpha \dot \rho = -\frac{m^2}{c} \rho \, .
\ea
This should be compared to the $\rho$ equation during the first inflationary phase where now $m^2 \rightarrow m^2/c$. This sudden change of mass induces a jump in $\ddot \rho$ which clearly can be seen in {\bf Fig.\ref{r-fig-chaotic} }.

Plugging solutions (\ref{rho-chaotic2}) and (\ref{rho-chaotic1}) into Eq. (\ref{Adot-chaotic})
one finds that $\dot A$ scales like $e^{  (4 c-1) \alpha}$ and $e^{3 \alpha}$
during the first and second inflationary stage respectively. The behavior of the gauge field during the first and second inflationary stages 
has a profile very similar to the plot on the left hand side of {\bf Fig.} \ref{sym-A-fig}. 
It is also instructive to look into the anisotropy parameter. During the first inflationary stage
one has
\ba
\delta \simeq \frac{2}{3} R_{1\, in} \exp \left[ 4 (c-1) \alpha 
\right] \, ,
\ea
whereas during the second inflationary stage it reaches the attractor value
\ba
\delta \simeq \frac{2 M_P^2}{3 \rho^2} \frac{c-1}{c^2} \simeq \frac{c-1}{3c} \epsilon \, .
\ea
As expected, $\delta \simeq \epsilon$ during the attractor regime.

Note that Eq. (\ref{rho-chaotic-eq2}) can be solved with the general answer 
\ba
\label{rho-chaotic-full}
\rho^2 \simeq -\frac{4 M_P^2 \alpha}{c} - \frac{M_P^2}{c} \ln \left[ \frac{4 (c-1)}{ K - e^{- 4 (c-1) \alpha} \left( K- 4 (c-1) e^{c \rho_{in}^2/M_P^2} \right)} \right] \, ,
\ea
where $K\equiv 4 c^2 p_A^2/m^2 M_P^2$. This has Eqs. (\ref{rho-chaotic1}) and (\ref{rho-chaotic2})
as the two limiting solutions. In {\bf Fig.} \ref{r-fig-chaotic} we have compared this analytical solution with the full numerical results and the agreement between them is very good. The time of the first phase change, $\alpha_1$, is when the second term in the denominator above becomes comparable to the first term which results in  
\ba
\label{alpha1-chaotic}
\alpha_1 \simeq  \frac{c \rho_{in}^2}{4(c-1)} + \frac{1}{4(c-1)} \ln \left[ \frac{m^2 M_P^2 (c-1)}{c^2 p_A^2} \right]
\simeq \frac{1}{4(c-1)} \ln \left[ \frac{c-1}{c^2} \frac{M_P^2}{\rho_{in}^2 R_{1\, in}}
\right] \, .
\ea
This indicates that the smaller is the value of the initial anisotropy $R_{1\, in}$, the longer it takes for the system to enter the attractor regime.  We have checked that this analytical expression gives a good estimate of $\alpha_1$.

\begin{figure}[t]
\includegraphics[width=7.5cm]{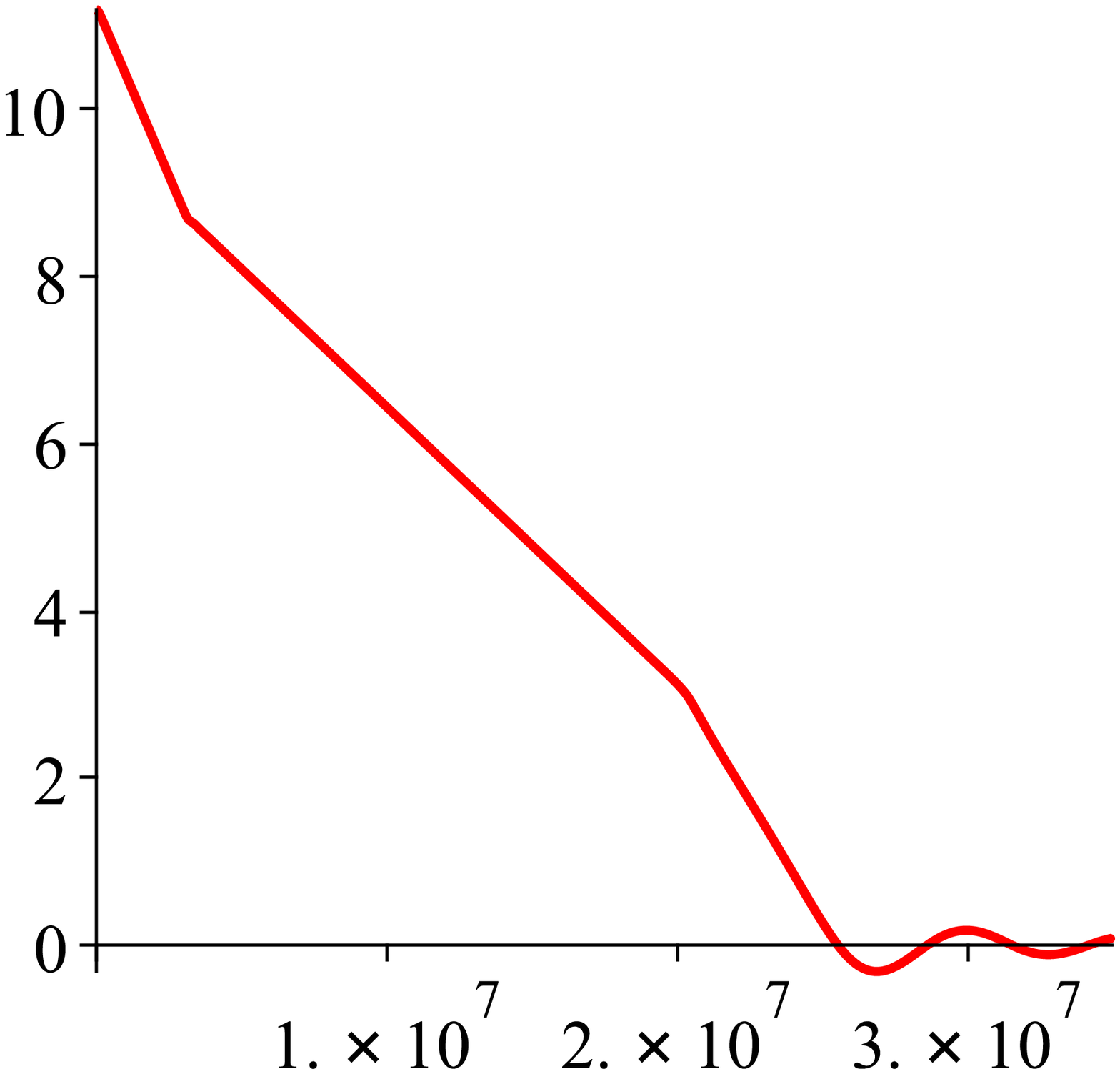} \hspace{0.5cm}
\includegraphics[width=7.5cm]{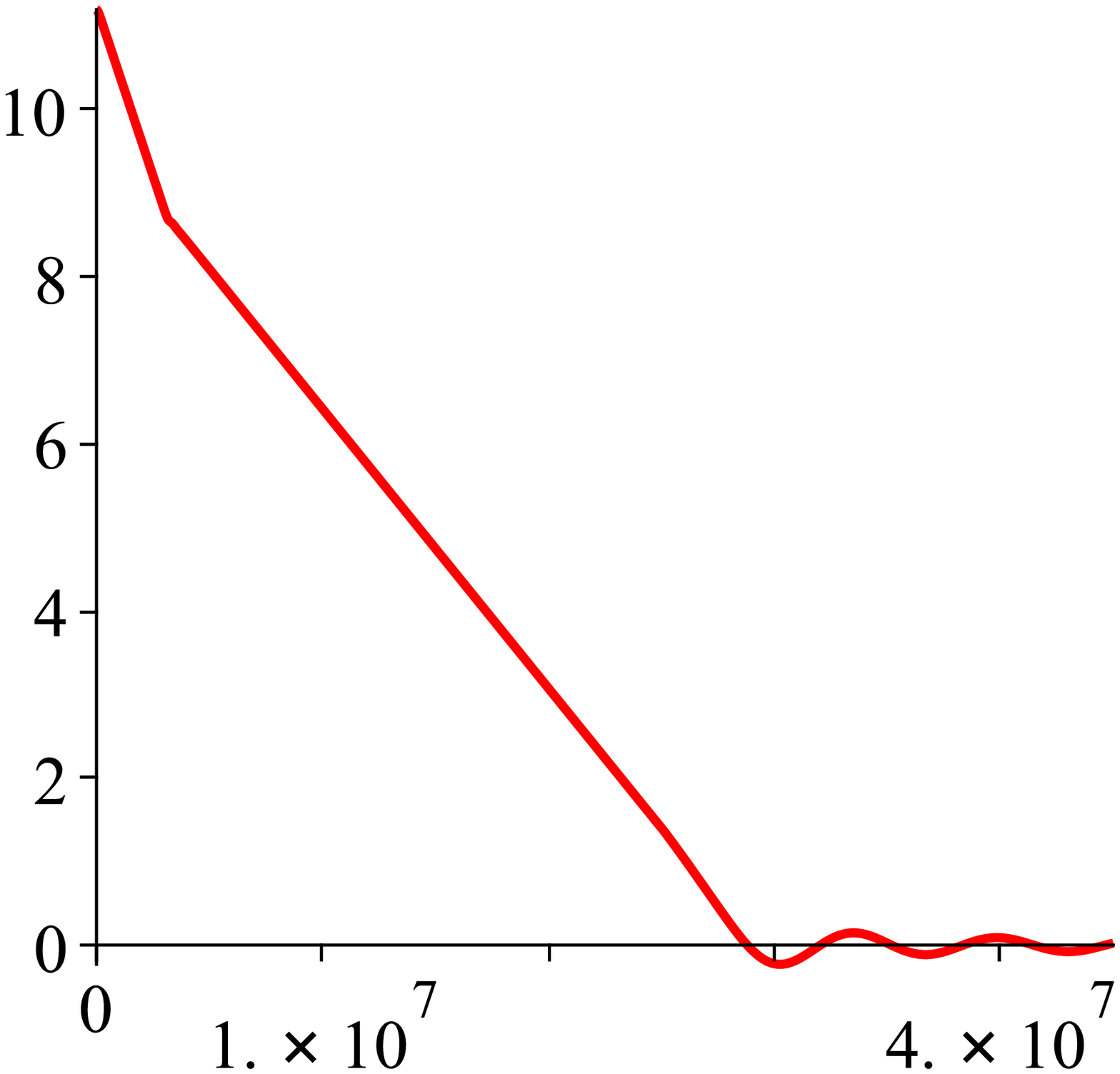} 
 \caption{ Here we present  the evolution of the inflaton field with respect to the time coordinate in chaotic inflation.  The left figure corresponds to $\e=0.1$ whereas for the right figure $\e=0$.  In this plot, the existence of the second kink and the duration of the third inflationary phase can be seen clearly when we turn on the gauge coupling $\e$. }
 \vspace{0.5cm}
\label{rho-time}
\end{figure}

Like in our previous examples, as the gauge field increases exponentially during the second inflationary stage the right hand side of Eq. (\ref{back-A-eq}) becomes important and one enters the final inflationary stage. The gauge field equation (in the slow-roll limit) has the same form as 
Eq. (\ref{A-eq1}) where now the parameter $\beta$ is given by
\ba
\beta \equiv \frac{6 e^2 M_P^4 (c-1)}{c^2 p_A^2}\, .
\ea
As in symmetry breaking case the solution is given by the Bessel function Eq. (\ref{A3-Bes}).
The start of the third inflationary stage, $\alpha=\alpha_2$, is when the argument of the Bessel function becomes comparable to unity so one obtains  $\alpha_2 \simeq\frac{-1}{4} \ln \beta $.
Our numerical analysis shows that this expression gives a good estimate of $\alpha_2$. As the argument of the Bessel function increases exponentially, the gauge field starts to oscillate rapidly. This in turn produces an oscillating effective mass for the inflaton in the form of
$e^2 \rho^2 A^2 e^{-2 \alpha}$ and the slow roll conditions are terminated quickly, ending inflation abruptly.  Our numerical analysis shows that usually inflation ends when the gauge field makes one or two oscillations in less than one e-fold.  The behavior of the gauge field during the final inflationary stage is similar to the plot on the right hand side of {\bf Fig.}
\ref{sym-A-fig} for the symmetry breaking potential. Also the behavior of the inflaton field as a function of time is presented in {\bf Fig.} \ref{rho-time}. The start of the third inflationary stage and the existence of the second kink can be seen clearly when we turn on
$\e$. However, as mentioned above, the third inflationary stage is very short, less than an e-fold. This can also be seen from {\bf Fig.} \ref{r-fig-chaotic}  where the evolution of the inflaton field is presented as a function of e-foldings.

\section{Conclusion and Discussions}
\label{conclusion}

In this work we studied anisotropic inflation in models with charge scalar fields. We have shown that, similar to \cite{Watanabe:2009ct}, the system reaches the attractor solutions sometime during inflation where the ratios $\delta/\epsilon$ and $R_1/\epsilon$, measuring the level of anisotropies, become at the order of unity. This attractor mechanism is fairly independent of the initial conditions. One can tune the model parameters such that the time of phase changes, denoted here by $\alpha_1$ and $\alpha_2$, take place within the first few e-foldings relevant for the CMB. 

The new interesting effect in our model is the effect of the gauge coupling $\e$  on the dynamics of the inflaton field and the gauge field. At the final stage of inflation, the term on the right hand side of Eq. (\ref{back-A-eq}) becomes important and gauge field become highly oscillatory.  Because of the interaction $\e^2 \rho^2 A^2$, the oscillations of the gauge field induce an effective time-dependent mass term for the inflaton field and inflation ends shortly after the gauge field starts to oscillate.  We have studied these effects in  examples of symmetry breaking and chaotic inflation  models. Both of our main results here, that is the existence of the attractor solutions and the oscillatory behavior of the gauge field at the end of inflation, show up  similarly in these two models. 

We also studied the effect of the gauge field on the dynamics of the  waterfall field in charged hybrid inflation model.  Because of the coupling $\e^2 \rho^2 A^2$ the onset of waterfall phase transition can be significantly different than in standard hybrid inflation. Furthermore, the highly oscillatory behavior of the gauge field and its coupling to the inflaton field can play important roles in the studies of tachyonic preheating and reheating. Also, as noticed in \cite{Yokoyama:2008xw}, the inhomogeneous end of inflation can have interesting effects for non-Gaussianities in this model.

In this work we considered only the background dynamics. It would be interesting to 
study the cosmological perturbations in our backgrounds and compare the results, such as the spectral index, the mixing between the scalar and tensor modes and non-Gaussianities, with observations.  We would like to come back to these questions in a future publication.
\vspace{1cm}

{\bf Note added:} while this work is completed, we became aware of the work  \cite{Jiro}
which has some overlaps with our results here. We thank the authors in \cite{Jiro} for letting us know about their work.

\section*{Acknowledgements}

We would like to thank K. Dimopoulos, S. Kanno, Y. Rodriguez, J. Soda,  J. P. Uzan and S. Yokoyama for useful discussions and correspondences. R. E. would like to thank 
D. Baumann, P. Creminelli,  M. Zaldarriaga for helpful discussions and ICTP for the hospitality during the ICTP summer school on cosmology, July 2010, when this work was in progress.  
H.F. would like to thank  Yukawa Institute for Theoretical Physics (YITP) for the hospitalities
during  the activities ``Gravity and Cosmology 2010" 
and ``YKIS2010 Symposium: Cosmology -- The Next Generation"   when this work was in progress.


\end{document}